\documentclass[aps,10pt,preprintnumbers,showkeys,eqsecnum,%
nofootinbib,showpacs]{revtex4-1}
\usepackage{bm} 
\usepackage{graphicx}
\usepackage{hyperref}
\newcommand{\be}{\begin{eqnarray}}
\newcommand{\ee}{\end{eqnarray}}
\newcommand{\beq}{\begin{equation}}
\newcommand{\eeq}{\end{equation}}
\begin{document}
\title{Light-front representation of chiral dynamics with
$\Delta$ isobar and large-$N_c$ relations}
\author{C.~Granados}
\email[E-mail: ]{cggranadosj@gmail.com}
\affiliation{Theory Center, Jefferson Lab, Newport News, VA 23606, USA}
\author{C.~Weiss}
\email[E-mail: ]{weiss@jlab.org}
\affiliation{Theory Center, Jefferson Lab, Newport News, VA 23606, USA}
\begin{abstract}
Transverse densities describe the spatial distribution of electromagnetic current in the nucleon
at fixed light-front time. At peripheral distances $b = O(M_\pi^{-1})$ the densities are governed 
by chiral dynamics and can be calculated model-independently using chiral effective field theory (EFT). 
Recent work has shown that the EFT results can be represented in first-quantized form, as overlap integrals 
of chiral light-front wave functions describing the transition of the nucleon to soft-pion-nucleon 
intermediate states, resulting in a quantum-mechanical picture of the peripheral transverse
densities. We now extend this representation to include intermediate states with $\Delta$ 
isobars and implement relations based on the large-$N_c$ limit of QCD. 
We derive the wave function overlap formulas for the $\Delta$ contributions to the peripheral transverse 
densities by way of a three-dimensional reduction of relativistic chiral EFT expressions. Our procedure 
effectively maintains rotational invariance and avoids the ambiguities with higher-spin particles 
in the light-front time-ordered approach. We study the interplay of $\pi N$ and $\pi \Delta$ intermediate 
states in the quantum-mechanical picture of the densities in a transversely polarized nucleon. 
We show that the correct $N_c$-scaling of the charge and magnetization densities emerges as the result of 
the particular combination of currents generated by intermediate states with degenerate $N$ and $\Delta$. 
The off-shell behavior of the chiral EFT is summarized in contact terms and can be studied easily. 
The methods developed here can be applied to other peripheral densities and to moments 
of the nucleon's generalized parton distributions.
\end{abstract}
\preprint{JLAB-THY-16-2235}
\maketitle
\tableofcontents
\newpage
\section{Introduction}
\label{sec:introduction}
Transverse densities are an essential concept in modern studies of hadron structure using 
theoretical and phenomenological methods \cite{Soper:1976jc,Burkardt:2000za,Burkardt:2002hr,Miller:2007uy}. 
They describe the distribution of physical quantities (charge, current, momentum) in the hadron at fixed 
light-front time $x^+ \equiv t + z$ and can be computed as the two-dimensional Fourier transforms of the 
invariant form factors associated with the operators. The transverse densities are frame-independent (invariant 
under longitudinal boosts) and enable an objective definition of the spatial structure of the hadron as a 
relativistic system. Their basic properties, connection with the parton picture (generalized parton 
distributions), and extraction from experimental data, have been discussed extensively in the 
literature; see Ref.~\cite{Miller:2010nz} for a review. In particular, the transverse densities offer 
a new way of identifying the chiral component 
of nucleon structure and studying its properties \cite{Strikman:2010pu,Granados:2013moa}. 
At transverse distances of the order of the inverse pion mass, $b = O(M_\pi^{-1})$, 
the densities are governed by chiral dynamics and can be computed 
model-independently using methods of effective 
field theory (EFT). The charge and magnetization densities in the nucleon's chiral periphery have been studied 
in a series of articles \cite{Granados:2013moa,Granados:2015rra,Granados:2015lxa}. 
The densities were computed in the relativistically invariant formulation of chiral 
EFT, using a dispersive representation that expresses them as integrals over the two-pion cut of the invariant 
form factors at unphysical momentum transfers $t > 4 M_\pi^2$ \cite{Granados:2013moa}. 
It was further shown that the chiral amplitudes producing the peripheral densities
can be represented as physical processes occurring in light-front time, involving the emission and absorption 
of a soft pion by the nucleon \cite{Granados:2015rra,Granados:2015lxa}. 
In this representation the peripheral transverse densities are expressed as overlap
integrals of chiral light-front wave functions, describing the transition of the original nucleon to a pion-nucleon 
state of spatial size $O(M_\pi^{-1})$ through the chiral EFT interactions. It enables a first-quantized, 
particle-based view of chiral dynamics, which reveals the role of pion orbital angular momentum 
and leads to a simple quantum-mechanical picture 
of peripheral nucleon structure.

Excitation of $\Delta$ isobars plays an important role in low-energy pion-nucleon dynamics. 
The $\Delta$ couples strongly to the $\pi N$ channel and can contribute to pionic processes, 
including those producing the peripheral transverse densities. Several extensions of chiral EFT 
with explicit isobars have been proposed, using different formulations for the spin-3/2 particle 
\cite{Hemmert:1997ye,Pascalutsa:2006up,CalleCordon:2012xz}.
Inclusion of the isobar is especially important in view of the large-$N_c$ limit of QCD, 
which implies general scaling relations for nucleon observables
\cite{'tHooft:1973jz,Witten:1979kh,Dashen:1993jt,Dashen:1994qi}. In the large-$N_c$ limit the $N$ and 
$\Delta$ become degenerate, $M_\Delta - M_N = O(N_c^{-1})$, and the $\pi NN$ and the $\pi N\Delta$ couplings 
are simply related, so that they have to be included on the same footing. Combination of intermediate 
$N$ and $\Delta$ contributions is generally needed for the chiral EFT results to satisfy the general 
$N_c$-scaling relations \cite{Cohen:1992uy,Cohen:1996zz}; see also Refs.~\cite{Strikman:2003gz,Strikman:2009bd}. 
The effect of the $\Delta$ on the peripheral transverse densities was studied in the dispersive 
representation in Ref.~\cite{Granados:2013moa}. It was observed that the densities obey the correct 
large-$N_c$ relations if $N$ and $\Delta$ contributions are combined; however, no mechanical explanation 
could be provided in this representation. 

In this article we develop the light-front representation of chiral dynamics with $\Delta$ isobars
and implement the large-$N_c$ relations in this framework. We introduce the light-front wave 
function of the $N \rightarrow \pi\Delta$ transition in chiral EFT in analogy to the $N \rightarrow \pi N$
one and derive the overlap representation of the peripheral transverse densities including both $\pi N$ 
and $\pi \Delta$ intermediate states. We then study the effects of the $\Delta$ in the resulting 
quantum-mechanical picture and explore the origin of the $N_c$-scaling of the transverse densities. 
The light-front representation with the isobar is of interest for several reasons. 
First, in the quantum-mechanical picture of the peripheral densities, the $\pi\Delta$ intermediate states 
allow for charge and rotational states of the peripheral pion different from those in $\pi N$ 
intermediate states, resulting in a rich structure. Second, the $N_c$-scaling of the peripheral 
densities can now be explained in simple mechanical terms, as the result of the cancellation (in the 
charge density) or addition (in the magnetization density) of the pionic convection current in the 
$\pi N$ and $\pi \Delta$ intermediate states. Third, the light-front representation can be extended
to study the $\Delta$ effects on the nucleon's peripheral partonic structure (generalized parton distributions),
including spin structure \cite{Strikman:2003gz,Strikman:2009bd}. 

We derive the light-front representation of chiral dynamics with $\Delta$ isobars by re-writing the 
relativistically invariant chiral EFT expressions for the current matrix element in a form that 
corresponds to light-front wave function overlap (three-dimensional reduction), generalizing 
the method developed for chiral EFT with the nucleon in Ref.~\cite{Granados:2015rra}. 
This approach offers several important advantages compared to light-front time-ordered perturbation
theory. First, it guarantees exact equivalence of
the light-front expressions with the relativistically invariant results. Second, it maintains 
rotational invariance, which is especially important in calculations with higher-spin particles
such as the $\Delta$, where breaking of rotational invariance would result in power-like divergences
of the EFT loop integrals. Third, it automatically generates the instantaneous (contact) terms
in the light-front formulation, which would otherwise have to be determined by extensive additional
considerations. The method of Ref.~\cite{Granados:2015rra} is thus uniquely suited to handle the isobar
in light-front chiral EFT and produces unambiguous results with minimal effort.
Similar techniques were used in studies of spatially integrated (non-peripheral) nucleon structure
in chiral EFT with nucleons in Refs.~\cite{Ji:2009jc,Burkardt:2012hk,Ji:2013bca}.

The extension of relativistic chiral EFT to spin-3/2 isobars is principally not unique. While the 
on-shell spinor of the spin-3/2 particle is unambiguously defined, the field theory requires off-shell 
extension of the projector in the propagator (Green function), which is inherently ambiguous
\cite{Pascalutsa:1999zz,Pascalutsa:2000kd}. The reason is that 
relativistic fields with spin $> 1/2$ involve unphysical degrees of freedom that have to be eliminated 
by constraints. In the EFT context this ambiguity is contained in the overall reparametrization invariance, 
which allows one to redefine the fields, changing the off-shell behavior of the propagator and
the vertices in a consistent fashion. Several schemes 
have been proposed for dealing with this ambiguity \cite{Hemmert:1997ye,Pascalutsa:2006up}. Fortunately, 
as will be shown below, the ambiguity in the field-theoretical formulation does not affect the light-front 
wave function overlap representation of the peripheral transverse densities in our approach. 
The differences caused by different off-shell extensions of the EFT reside entirely in contact (instantaneous) 
terms in the densities and can easily be quantified. We perform our calculations using the relativistic 
Rarita-Schwinger formulation for the $\Delta$ propagator and an empirical $\pi N \Delta$ coupling,
but otherwise no explicit off-shell terms at the Lagrangian level. We justify this choice by showing
that the sum of $N$ and $\Delta$ contributions obeys the correct $N_c$-scaling (in both the 
wave-function overlap and the contact terms), so that no explicit off-shell terms are required
from this perspective. This level of treatment is sufficient for our purposes. Going beyond it would
require a dynamical criterion to fix the off-shell ambiguity of the EFT.

In our approach to chiral dynamics we consider nucleon structure at ``chiral'' distances $b = O(M_\pi^{-1})$,
where we regard the pion mass as parametrically small compared to the ``non-chiral'' inverse hadronic size 
but do not consider the limit $M_\pi \rightarrow 0$ \cite{Granados:2013moa}.
When introducing the $\Delta$ isobar in this scheme we do not assume any parametric relation between the 
$N$-$\Delta$ mass splitting and the pion mass. The large-$N_c$ limit is then implemented by assuming the
standard scaling relations $M_\Delta - M_N = O(N_c^{-1})$ and $M_\pi = O(N_c^0)$, which 
implies that the region of chiral distances remains stable in the large-$N_c$ limit,
$b = O(M_\pi^{-1}) = O(N_c^0)$, as is physically sensible. From a dynamical standpoint, the introduction
of the isobar into the chiral EFT changes only the coupling of the peripheral pions to the nucleon, 
not their propagation to distances $b = O(M_\pi^{-1})$, which defines the basic character of the 
peripheral densities. An advantage of the spatial identification of the chiral region is precisely that 
it allows us to introduce the isobar in this natural manner.

The plan of the article is as follows. In Sec.~\ref{sec:densities} we review the basic features
of the transverse densities and their behavior in the chiral periphery. In Sec.~\ref{sec:chiral} we summarize 
the description of the isobar in chiral EFT (spinors, propagator, vertices) as relevant for the
present calculation. In Sec.~\ref{sec:lightfront} we derive the light-front representation
of the peripheral transverse densities in chiral EFT. This includes the 3-dimensional reduction
of the relativistically invariant chiral EFT expressions, the wave function overlap representation, 
the light-front spin states of $N$ and $\Delta$ baryons, the coordinate-space light-front wave functions,
and the properties of the transverse densities with $N$ and $\Delta$ intermediate states.
In Sec.~\ref{sec:mechanical} we discuss the wave function overlap representation with transversely 
polarized spin states and the resulting quantum-mechanical picture. In Sec.~\ref{sec:largenc} we
implement the large-$N_c$ relations between the EFT parameters and study the scaling behavior of the 
chiral light-front wave functions and peripheral transverse densities; we also evaluate the contact term 
contributions and discuss the role of the off-shell ambiguity in the light-front representation. 
A summary and outlook are presented in Sec.~\ref{sec:summary}.

The study reported here draws extensively on our previous work \cite{Granados:2013moa,Granados:2015rra}.
Because the intermediate $\Delta$ contributions in the light-front representation 
need to be discussed together with the intermediate $N$ 
ones calculated in Ref.~\cite{Granados:2015rra}, some repetition of arguments and quoting of the nucleon 
expressions is unavoidable to make the presentation readable. Generally, in each step of the derivations
in Secs.~\ref{sec:lightfront} and \ref{sec:mechanical}, we summarize the basic ideas as formulated 
in Refs.\cite{Granados:2013moa,Granados:2015rra}, give the expressions for the intermediate $N$, 
and present the explicit calculations for the $\Delta$.
\section{Transverse densities}
\label{sec:densities}
\subsection{Basic properties}
%
%
\begin{figure}[t]
\begin{center}
\includegraphics[width=.62\textwidth]{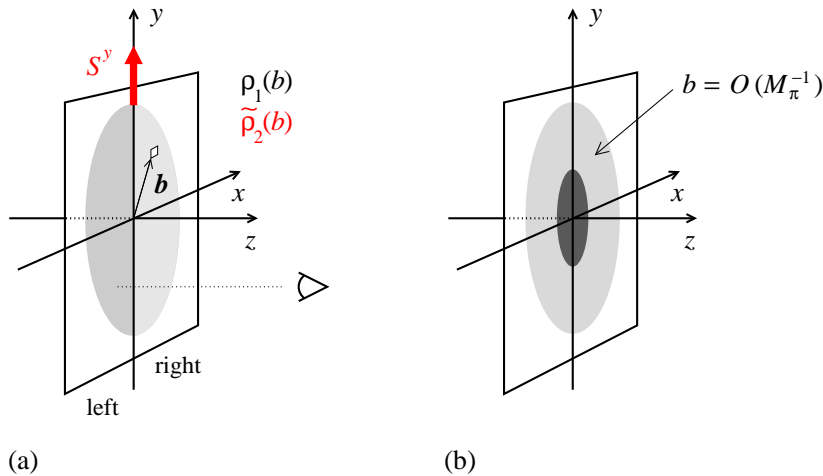}
\end{center}
\caption[]{(a) Interpretation of the transverse densities of the 
electromagnetic current in a nucleon state with 
spin quantized in the transverse $y$-direction, Eq.~(\ref{j_plus_rho}). 
$\rho_1(b)$ describes the spin-independent (or left-right symmetric) 
part of the plus current density; $\cos\phi \, \widetilde\rho_2(b)$ 
describes the spin-dependent (or left-right asymmetric) part. (b) Peripheral region
dominated by chiral dynamics. The central area represents the nucleon's non-chiral hadronic size.}
\label{fig:densities}
\end{figure}
The matrix element of the electromagnetic current between nucleon states with
4-momenta $p_{1, 2}$ and spin quantum numbers $\sigma_{1, 2}$ is of 
the general form (we follow the conventions of Ref.~\cite{Granados:2013moa})
\beq
\langle N(p_2, \sigma_2) | J^\mu (x) | N(p_1, \sigma_1) \rangle\
\;\; = \;\; \bar u_2 \left[ \gamma^\mu F_1(t) - 
\frac{\sigma^{\mu\nu} \Delta_\nu}{2 M_N} F_2(t) \right] u_1 \, 
e^{i\Delta x} ,
\label{me_general}
\eeq
where $u_1 \equiv u(p_1, \sigma_1)$ \textit{etc.}\ are the nucleon bispinors, normalized to 
$\bar u_1 u_1 = \bar u_2 u_2 = 2 M_N$; $\sigma^{\mu\nu} \equiv \frac{1}{2}\left[ \gamma^\mu, \gamma^\nu\right]$,
and 
\beq
\Delta \;\; \equiv \;\; p_2 - p_1 
\eeq
is the 4-momentum transfer. The dependence on the space-time point $x$, where the current is measured,
is dictated by translational invariance. The Dirac and Pauli form factors, $F_1$ and $F_2$, are functions 
of the Lorentz-invariant momentum transfer
\beq
t \;\; \equiv \;\; \Delta^2 \;\; = \;\; (p_2 - p_1)^2,
\eeq
with $t < 0$ in the physical region for electromagnetic scattering. While the physical content of the invariant 
form factors can be discussed independently of any reference frame, their space-time interpretation
depends on the formulation of relativistic dynamics and on the reference frame.

In the light-front formulation of relativistic dynamics one considers the evolution of 
strong interactions in light-front time $x^+ \equiv x^0 + x^3 = x^0 + z$, as corresponds 
to clocks synchronized by a light wave traveling through the system in the $z$-direction
\cite{Dirac:1949cp,Leutwyler:1977vy,Brodsky:1997de}.
Particle states are characterized by their light-front momentum $p^+ \equiv p^0 + p^z$ and 
transverse momentum $\bm{p}_T \equiv (p^x, p^y)$, while $p^- \equiv p^0 - p^z = (p_T^2 + M_N^2)/p^+$
plays the role of the energy. In this context it is natural to consider the current matrix element 
Eq.~(\ref{me_general}) in a class of reference frames where the momentum transfer is in the 
transverse direction
\beq
\Delta^\pm \; = \; 0, \hspace{2em}
\bm{\Delta}_T \; = \; \bm{p}_{2T} - \bm{p}_{1T} \; \neq \; 0 ,
\label{Delta_transverse}
\eeq
and to represent the form factors as Fourier transforms of certain transverse densities
\cite{Miller:2007uy,Miller:2010nz}
\beq
F_{1, 2}(t = -\bm{\Delta}_T^2) \;\; = \;\; \int d^2 b \; 
e^{i \bm{\Delta}_T \cdot \bm{b}} \; \rho_{1, 2} (b) ,
\label{rho_def}
\eeq
where $\bm{b} \equiv (b^x, b^y)$ is a 2-dimensional coordinate variable
and $b \equiv |\bm{b}|$. The basic properties of the transverse densities 
$\rho_{1, 2}(b)$ and their physical interpretation have been discussed
in the literature \cite{Soper:1976jc,Burkardt:2000za,Burkardt:2002hr,Miller:2007uy,Miller:2010nz} 
and are summarized in Ref.~\cite{Granados:2013moa}.
They describe the transverse spatial distribution of the light-front 
plus current, $J^+ \equiv J^0 + J^z$, in the nucleon at fixed light-front time $x^+$. 
A simple interpretation emerges if the nucleon spin states are chosen
as light-front helicity states, which are constructed by preparing rest-frame
spin states and boosting them to the desired light-front plus and transverse
momentum \cite{Brodsky:1997de,Granados:2013moa} (see also Sec.~\ref{subsec:polarization} below). 
In a state where the nucleon is localized in transverse space at the origin,
and polarized along the $y$-direction in the rest frame,
the matrix element of the current $J^+(x)$ at light-front time
$x^+ = 0$ and light-front coordinates $x^- = 0$ and 
$\bm{x}_T = \bm{b}$ is given by
\be
\langle J^+ (\bm{b}) \rangle_{\text{\scriptsize localized}}
&=& (...) \; \left[
\rho_1 (b) \;\; + \;\; (2 S^y) \, \cos\phi \, \widetilde\rho_2 (b) \right] ,
\label{j_plus_rho}
\\[2ex]
\widetilde\rho_2 (b) &\equiv& \frac{\partial}{\partial b} 
\left[ \frac{\rho_2(b)}{2 M_N} \right] ,
\label{rho_2_tilde_def}
\ee
where $(...)$ stands for a kinematic factor arising from the normalization of the localized
nucleon states \cite{Granados:2013moa}, $\cos\phi \equiv b^x/b$ is the cosine of the angle
of $\bm{b}$ relative to the $x$-axis, and $S^y = \pm 1/2$ is the spin projection on the $y$-axis
in the nucleon rest frame. One sees that $\rho_1(b)$ describes the spin-independent 
part of the current, while the function 
$\cos\phi \, \widetilde\rho_2 (b)$ describes the spin-dependent part of the current in the
transversely polarized nucleon (see Fig.~\ref{fig:densities}a). 

An alternative interpretation appears if one considers the localized nucleon state with
spin $S^y = + 1/2$ and takes the current at two opposite points on the $x$-axis, 
$\bm{b} = \mp b \bm{e}_x$, i.e., on the ``left'' and ``right'' when looking at the nucleon
from $z = +\infty$ (see Fig.~\ref{fig:densities}a). 
According to Eq.~(\ref{j_plus_rho}) the current at these points is given by
\beq
\langle J^+ (\mp b\bm{e}_x) \rangle_{\text{\scriptsize localized}}
\;\; = \;\; (...) \; [ \rho_1 (b) \, \mp \, \widetilde\rho_2 (b) ]
\;\; \equiv \;\;  (...) \; \rho_{\rm left/right}(b) ,
\label{rho_left_right}
\eeq
such that
\beq
\left.
\begin{array}{l}
\rho_1 (b) \\[1ex] \widetilde{\rho}_2 (b) 
\end{array}
\right\}
\;\; = \;\; \frac{1}{2} [ \pm \rho_{\rm left}(b) + \rho_{\rm right}(b)] .
\label{rho_1_2_from_left_right}
\eeq
One sees that $\rho_1(b)$ and $\widetilde\rho_2(b)$ can be interpreted as 
the left-right symmetric and antisymmetric parts of the plus current on the $x$-axis.
This representation is useful because the left and right transverse densities have
simple properties in dynamical models (positivity, mechanical interpretation).
We shall refer to it extensively in the present study \cite{Granados:2015rra,Granados:2015lxa}.

\subsection{Chiral periphery}
At distances $b = O(M_\pi^{-1})$ the nucleon's transverse densities are governed by the effective
dynamics resulting from the spontaneous breaking of chiral symmetry (see Fig.~\ref{fig:densities}b)
\cite{Strikman:2010pu,Granados:2013moa}. The transverse densities in this region arise from
chiral processes in which the current couples to the nucleon through $t$-channel exchange 
of soft pions. In the densities $\rho_1$ and $\widetilde\rho_2$ associated with the
electromagnetic current, Eq.~(\ref{me_general}), the leading large-distance 
component results from the exchange of two pions and occurs in the isovector densities
\beq
\rho_1^V \equiv {\textstyle\frac{1}{2}}(\rho_1^p - \rho_1^n) , \hspace{2em}
\widetilde\rho_2^V \equiv {\textstyle\frac{1}{2}}(\widetilde\rho_2^p - \widetilde\rho_2^n) .
\eeq
The large-distance behavior is of the form
\beq
\left.
\begin{array}{lcl}
\rho_1^V(b) &\sim& P_1(b) \\[1ex]
\widetilde\rho_2^V(b) &\sim& \widetilde P_2(b) 
\end{array} \right\} \; \times \exp(-2M_\pi b) .
\label{large_b_general}
\eeq
The exponential decay is governed by the minimal mass of the exchanged system, $2 M_\pi$.
The pre-exponential factors $P_1$ and $\widetilde{P}_2$ are determined by the coupling of 
the exchanged system to the nucleon and exhibit a rich structure, as their dependence on
$b$ is governed by multiple dynamical scales: $M_\pi, M_N$, and possibly also $M_\Delta - M_N$. 

%
%
\begin{figure}[t]
\includegraphics[width=.4\textwidth]{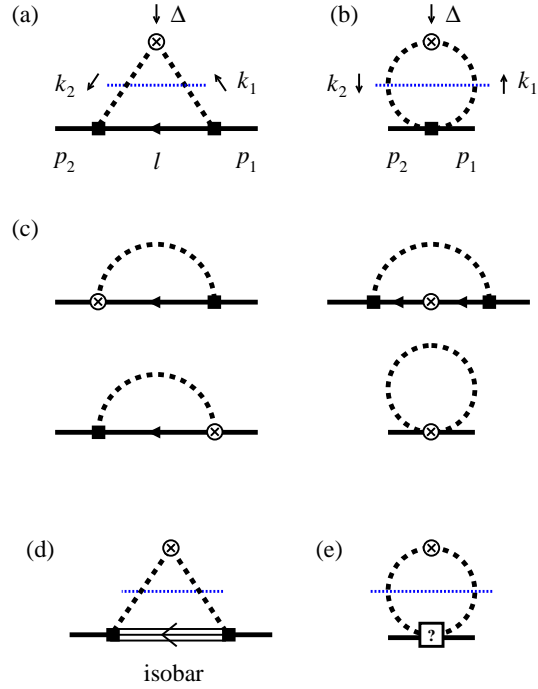}
\caption[]{Feynman diagrams of LO chiral EFT processes contributing 
to the the electromagnetic current matrix element. (a,b)~$\pi N$ diagrams contributing 
to the two-pion cut (indicated by the dashed blue line), or the peripheral transverse densities. 
(c)~Diagrams not contributing to the two-pion cut,or the peripheral transverse densities. 
(d, e)~$\pi\Delta$ diagrams contributing to the two-pion cut. Diagram (e) would result from 
a ``new'' $\pi\pi N N$ contact coupling accompanying the introduction of the $\Delta$ 
(see Sec.~\ref{subsec:couplings}).}
\label{fig:diag}
\end{figure}
In the chiral region $b = O(M_\pi^{-1})$ the transverse densities can be calculated using 
chiral EFT \cite{Strikman:2010pu,Granados:2013moa,Granados:2015rra,Granados:2015lxa}.
The densities can be obtained from the known chiral EFT results for the invariant 
form factors \cite{Gasser:1987rb,Bernard:1992qa,Kubis:2000zd,Kaiser:2003qp}. 
It is necessary to use chiral EFT with relativistic nucleons \cite{Becher:1999he} in order
to reproduce the exact analytic structure of the form factors near the two-pion threshold 
at $t = 4 M_\pi^2$ (including a sub-threshold singularity on the unphysical sheet), as the
latter determines the basic large-distances behavior of the densities, Eq.~(\ref{large_b_general}).
In the LO approximation the processes contributing to 
the peripheral densities $\rho_1^V(b)$ and $\widetilde\rho_2^V (b)$ are given by the Feynman 
diagrams of Fig.~\ref{fig:diag}a and b, in which the current couples to the nucleon through
two-pion exchange in the $t$-channel (explicit expressions will be given below). 
These processes contribute to the two-pion cut of the form factor and produce densities of the 
form Eq.~(\ref{large_b_general}) \cite{Strikman:2010pu,Granados:2013moa}. Diagrams of 
Fig.~\ref{fig:diag}c, in which the current couples directly to the nucleon, do not contribute
to the two-pion cut of the form factor and produce densities with range $O(M_N^{-1})$ 
or terms $\propto \delta^{(2)}(\bm b)$. These diagrams renormalize the current
at the center of the nucleon and do not need to be considered in the calculation of peripheral
densities at $b = O(M_\pi^{-1})$.

Excitation of $\Delta$ isobars contributes to the peripheral transverse densities 
by modifying the coupling of the two-pion exchange to the nucleon and can be 
included in the chiral EFT calculations \cite{Strikman:2010pu,Granados:2013moa}. 
In a relativistic formulation at LO level the relevant processes are those described by the
Feynman diagrams of Fig.~\ref{fig:diag}d and e (explicit expressions will be given below).
Diagram Fig.~\ref{fig:diag}d describes the excitation of a dynamical $\Delta$ through the
$\pi N\Delta$ coupling; diagram Fig.~\ref{fig:diag}e represents possible ``new'' contact 
couplings associated with the introduction of the $\Delta$ into the chiral EFT.

The LO chiral EFT results for the peripheral densities can be represented in two equivalent forms.
In the dispersive representation \cite{Strikman:2010pu,Granados:2013moa} the densities 
are expressed as integrals over the imaginary 
parts of the form factor, $\textrm{Im}\, F_{1, 2}(t)/\pi$, on the two-pion cut at $t > 4 M_\pi^2$
(spectral functions). This representation is convenient for studying the asymptotic behavior of 
the densities, which is related to the behavior of the spectral function near threshold $t \rightarrow 4 M_\pi^2$.
In the light-front representation \cite{Granados:2015rra,Granados:2015lxa} 
the densities are expressed as overlap integrals of chiral light-front 
wave functions, describing the transition of the external nucleon to a pion-nucleon intermediate state
of size $O(M_\pi^{-1})$ through the chiral EFT interactions, as well as contact terms.
It enables a first-quantized view of the chiral processes and 
leads to a simple quantum-mechanical picture of peripheral nucleon structure. 
Our aim in the following is to derive the light-front representation of the $\Delta$ contributions. 
\section{Chiral dynamics with $\Delta$ isobar}
\label{sec:chiral}
\subsection{Spinors and propagator}
\label{subsec:spinors}
We first want to summarize the elements of the field-theoretical description of the $\Delta$ and 
its coupling to pions, as needed for the derivation of the light-front representation.
The nucleon in relativistically invariant chiral EFT is described by a relativistic 
spin-1/2 field (Dirac field). The free nucleon state with 4-momentum $l$, subject to the
mass-shell condition $l^2 = M_N^2$, is described by a bispinor $u(l, \sigma)$ with 
spin quantum number $\sigma = \pm 1/2$, satisfying the Dirac equation ($\hat l \equiv l^\mu \gamma_\mu$)
\beq
(\hat l - M_N) u(l, \sigma) \;\; = \;\; 0,
\label{dirac_nucleon}
\eeq
and normalized such that
\beq
\bar u (l, \sigma_2 ) u(l, \sigma_1 ) \;\; = \;\; 2 M_N \, \delta(\sigma_2, \sigma_1).
\label{normalization_nucleon}
\eeq
The specific choice of spin polarization in our calculations will be described below 
(see Sec.~\ref{subsec:polarization}). The propagator of the spin-1/2 field with 
causal boundary condition (Feynman propagator) is
\beq
G(l) \;\; \equiv \;\; \frac{\hat l + M_N}{l^2 - M_N^2 + i0} ,
\eeq
where the 4-momentum is generally off mass-shell, $l^2 \neq M_N^2$, and the residue
at the pole at $l^2 = M_N^2$ is the projector on nucleon states,
\beq
(\hat l + M_N) |_{l^2 = M_N^2} \;\; = \;\; \sum_{\sigma} u(l, \sigma) \bar u(l, \sigma) .
\label{projector_nucleon}
\eeq

The isobar in relativistic EFT is described by a relativistic spin-3/2 field. The general procedure
is to construct this field as the product of a spin-1 (vector) and a spin-1/2 (spinor) field and 
eliminate the spin-1/2 degrees of freedom through constraints formulated in a
relativistically covariant manner (Rarita-Schwinger formulation); see Ref.~\cite{Hemmert:1997ye}
for a review. The free isobar state with 4-momentum $l$, subject to the mass-shell condition 
$l^2 = M_\Delta^2$, is described by a vector-bispinor
\be
U^\mu(l, \sigma) &\equiv& \sum_{\lambda \sigma'} C^{3/2, \sigma}_{1, \lambda;\; 1/2, \sigma'} \;
\epsilon^\mu (l, \lambda) \; u (l, \sigma') ,
\label{vector_bispinor}
\ee
with spin quantum number $\sigma = (-3/2, -1/2, 1/2, 3/2)$. Here
\be
C^{3/2, \sigma}_{1, \lambda;\; 1/2, \sigma'} &\equiv& 
\langle 1 \lambda, {\textstyle\frac{1}{2}} \sigma'| {\textstyle\frac{3}{2}}, \sigma \rangle
\ee
is the vector coupling coefficient for angular momenta $1 + \frac{1}{2} \rightarrow \frac{3}{2}$;
$\epsilon^\mu(l, \lambda)$ is the polarization vector of a spin-1 particle
with 4-momentum $l$ and spin quantum number $\lambda = (-1, 0, 1)$ and satisfies
\be
l^\mu \epsilon_\mu (l, \lambda) &=&  0 ,
\label{polarization_vector_constraint}
\\[1ex]
\sum_{\lambda} \epsilon^\mu (l, \lambda) \epsilon^\nu (l, \lambda) &=&  
- g^{\mu\nu} + \frac{l^\mu l^\nu}{M_\Delta^2} ;
\label{polarization_vector_completeness}
\ee
and $u (l, \sigma')$ is a bispinor with spin quantum number $\sigma' = \pm 1/2$ 
and mass $M_\Delta$ and satisfies [cf.~Eqs.~(\ref{dirac_nucleon})
and (\ref{normalization_nucleon})]
\be
(\hat l - M_\Delta ) u (l, \sigma') &=& 0, 
\label{spinor_delta_dirac}
\\[2ex]
\bar u (l, \sigma^{\prime\prime}) u(l, \sigma') &=& 2 M_\Delta \, \delta(\sigma^{\prime\prime}, \sigma'),
\label{spinor_delta_normalization}
\\[1ex]
(\hat l + M_\Delta) |_{l^2 = M_N^2} &=& \sum_{\sigma'} u(l, \sigma') \bar u(l, \sigma') .
\label{spinor_delta_projector}
\ee
As a result the vector-bispinor Eq.~(\ref{vector_bispinor}) satisfies the conditions
\be
(\hat l - M_\Delta) U^\mu &=& 0,
\label{vector_bispinor_dirac}
\\[1ex]
l^\mu U_\mu &=& 0.
\label{vector_bispinor_l}
\ee
The constraint Eq.~(\ref{polarization_vector_constraint}) eliminates spin-0 states
and ensures that the polarization vector describes only spin-1 states in the rest frame. 
Equation~(\ref{vector_bispinor_l}) therefore implies that the vector-bispinor contains
only spin-3/2 degrees of freedom in the rest frame. 
The normalization of the vector-bispinor is such that
\beq
\bar U^\mu (l, \sigma_2 ) U_\mu (l, \sigma_1 ) \;\; = \;\; - 2 M_\Delta \, \delta(\sigma_2, \sigma_1) .
\eeq

The propagator of the relativistic spin-3/2 field is of the general form
\beq
G^{\mu\nu}(l) \;\; \equiv \;\; \frac{R^{\mu\nu}(l)}{l^2 - M_\Delta^2 + i0} ,
\label{green_delta}
\eeq
where $R^{\mu\nu}(l)$ represents a 4-tensor and matrix in bispinor indices and is defined for 
arbitrary 4-momenta $l$, not necessarily restricted to the mass shell. 
On the mass shell $l^2 = M_\Delta^2$, i.e., at the pole
of the propagator, $R^{\mu\nu}(l)$ coincides with the projector on isobar states
[cf.\ Eq.~(\ref{projector_nucleon})]
\beq
R^{\mu\nu}(l) |_{l^2 = M_\Delta^2} \;\; = \;\; \sum_{\sigma} U^\mu (l, \sigma)
\bar U^\nu (l, \sigma) .
\label{projector_delta}
\eeq
On the mass shell it therefore satisfies the constraints [cf.\ Eqs.~(\ref{vector_bispinor_dirac})
and (\ref{vector_bispinor_l})]
\beq
\left.
\begin{array}{r}
l_\mu R^{\mu\nu}  \\[1ex]
R^{\mu\nu} l_\nu   \\[1ex]
(\hat l - M_\Delta) R^{\mu\nu} \\[1ex]
R^{\mu\nu} (\hat l - M_\Delta)
\end{array}
\right\}
\;\; = \;\; 0 \hspace{2em} \textrm{at} \hspace{2em} l^2 = M_\Delta^2 .
\label{projector_constraint}
\eeq
An explicit representation of $R^{\mu\nu}(l)$ on the mass shell is
\be
R^{\mu\nu} (l)|_{l^2 = M_\Delta^2} &=& (\hat l + M_\Delta )
\left[ -g^{\mu\nu} + \frac{1}{3} \gamma^\mu \gamma^\nu
+ \frac{2}{3 M_\Delta^2} l^\mu l^\nu
- \frac{1}{3 M_\Delta} (l^\mu \gamma^\nu - \gamma^\mu l^\nu )
\right]
\label{projector_expanded}
\\
&=& (\hat l + M_\Delta) \left( -g^{\mu\nu} + 
\frac{l^\mu l^\nu}{M_\Delta^2}
\right)
\; - \; 
\frac{1}{3} 
\left( \gamma^\mu + \frac{l^\mu}{M_\Delta} \right)
(\hat l - M_\Delta)
\left( \gamma^\nu + \frac{l^\nu}{M_\Delta} \right) .
\label{projector_factorized}
\ee

The extension of the function $R^{\mu\nu} (l)$ to off mass-shell momenta $l^2 \neq M_\Delta^2$ is inherently
not unique \cite{Pascalutsa:1999zz,Pascalutsa:2000kd}. 
The reason is that the physical constraints eliminating the spin-1/2 content of the field can
be formulated unambiguously only for on-shell momenta. Any tensor-bispinor matrix function $R^{\mu\nu}(l)$ 
that satisfies the conditions Eq.~(\ref{projector_constraint}) on-shell represents a valid propagator 
Eq.~(\ref{green_delta}). In EFT this ambiguity in the definition of the propagator is accompanied by a similar 
ambiguity in the definition of the vertices and contained in the overall reparametrization invariance of the theory,
i.e., the freedom of redefining the fields off mass-shell \cite{Hemmert:1997ye}. For our purposes --- deriving
the light-front wave function representation of the isobar contributions to the transverse densities --- 
this ambiguity will be largely unimportant, as it shows up only in contact terms where its effects can easily 
be quantified. We can therefore perform our calculation with a specific choice of the spin-3/2 
propagator. \textit{We define} the spin-3/2 propagator by using in Eq.~(\ref{green_delta}) the expression for
$R^{\mu\nu} (l)$ given by the off-shell extension of the expressions in Eq.~(\ref{projector_expanded}) 
or (\ref{projector_factorized}) (note that the two expressions are algebraically equivalent also for 
off-shell momenta). This choice represents the natural generalization of the projector to off-shell
momenta and leads to simple expressions for the isobar contributions to the transverse densities
and other nucleon observables. We shall justify this choice a posteriori by showing that the resulting 
isobar contributions obey the correct large-$N_c$ relations when combined with the nucleon contributions
(see Sec.~\ref{subsec:contact}).
\subsection{Coupling to pion}
\label{subsec:couplings}
The interaction of the nucleon with pions are constrained by chiral invariance. The explicit vertices
emerge from the expansion of the non-linear chiral Lagrangian in pion field \cite{Becher:1999he}.
The vertices governing the LO chiral processes in the diagrams of Fig.~\ref{fig:diag}a and b 
are contained in
\be
\mathcal{L}_{\pi N N} &=& 
- \frac{g_A}{2 F_\pi} \bar\psi_j \gamma^\mu \gamma_5 \, (\tau^a)^j_i \, \psi^i 
\; \partial_\mu \pi^a 
\; - \; \frac{1}{4F_\pi^2} \bar\psi_j \gamma^\mu \, (\tau^a)^j_i \, \psi^i \;
\epsilon^{abc} \pi^b \partial_\mu \pi^c ,
\label{lag_n}
\ee
where $\psi^i = (\psi^p, \psi^n)$ is the isospin-1/2 nucleon field, $\pi^a = (\pi^1, \pi^2, \pi^3)$ are
the real (cartesian) components of the isospin-1 pion field, and summation over isospin indices is implied.
The first term in Eq.~(\ref{lag_n}) describes 
a three-point $\pi NN$ coupling. The axial vector vertex is equivalent to the 
conventional pseudoscalar vertex under the conditions that (i) the nucleons are on-shell;
(ii) 4-momenta at the vertex are conserved, i.e., the pion 4-momentum is equal to the difference of the 
nucleon 4-momenta, as is the case in Feynman perturbation theory. The relation is:
\beq
- \frac{g_A}{2 F_\pi} 
\bar u(l') i (\hat l' - \hat l) \gamma_5 \tau^a u(l)
\;\; = \;\; 
\frac{g_A M_N}{F_\pi} 
\bar u(l') i \gamma_5 \tau^a u (l)
\;\; \equiv \;\; g_{\pi NN} \bar u(l') i \gamma_5 \tau^a u(l) .
\label{axial_pseudoscalar}
\eeq
In terms of the complex (spherical) components of the pion field the first term in Eq.~(\ref{lag_n}) reads
\beq
 - \frac{g_A}{2 F_\pi} \left( 
\sqrt{2} \, \bar\psi_n \gamma^\mu \gamma_5 \partial_\mu \pi^+ \psi_p 
\; + \; \sqrt{2} \, \bar\psi_p \gamma^\mu \gamma_5 \partial_\mu \pi^+ \psi_n
\; + \; \bar\psi_p \gamma^\mu \gamma_5 \partial_\mu \pi^0 \psi_p 
\; - \; \bar\psi_n \gamma^\mu \gamma_5 \partial_\mu \pi^0 \psi_n 
\right) .
\eeq
The second term in Eq.~(\ref{lag_n}) represents a four-point $\pi\pi N N$ coupling. It arises due to
chiral invariance, as can be seen in the fact that its coupling is completely given in terms of $F_\pi$ 
and does not involve any dynamical parameters. 
In LO the parameters are taken at their physical 
values $g_A = 1.26$ and $F_\pi = 93\, \textrm{MeV}$.

The coupling of the $\Delta$ isobar to the $\pi N$ system is likewise constrained by chiral 
invariance \cite{Hemmert:1997ye}. The unique structure emerging as the first-order expansion in 
the pion field of the nonlinear chirally invariant $\pi N \Delta$ Lagrangian is 
\beq
\mathcal{L}_{\pi N \Delta} \;\; \sim \;\; \frac{i g_{\pi N \Delta}}{M_N} \;
\; C^{3/2, j}_{1, k;\; 1/2, i} \;
\bar\psi^i \; \partial_\mu \pi^k \; \Psi^{\mu j} \; + \; \mbox{h.c.},
\label{lag_delta_general}
\eeq
where $\Psi_{\Delta}^{\mu j} = (\Psi_{\Delta ++}^{\mu}, \Psi_{\Delta +}^{\mu}, \Psi_{\Delta 0}^{\mu}, 
\Psi_{\Delta -}^{\mu})$ is the isobar field, $\pi^k = (\pi^+, \pi^-, \pi^0)$ are 
the complex (spherical) components of the pion field, and $C^{3/2, i}_{1, k;\; 1/2, j}$ denotes 
the isospin vector coupling coefficient in the appropriate representation. In terms of the 
individual isospin components the couplings are
\be
\mathcal{L}_{\pi N \Delta} &=& \frac{i g_{\pi N \Delta}}{\sqrt{2} M_N}
\left( 
\bar\psi_p \; \partial_\mu \pi^- \; \Psi_{\Delta ++}^\mu 
+ \sqrt{\frac{2}{3}} \; \bar\psi_p \; \partial_\mu \pi^0 \; 
\Psi_{\Delta +}^\mu
+ \frac{1}{\sqrt{3}} \; \bar\psi_p \; 
\partial_\mu \pi^+ \; \Psi_{\Delta 0}^\mu
\right. 
\nonumber
\\
&& + \; 
\left.
\bar\psi_n \; \partial_\mu \pi^+ \; \Psi_{\Delta -}^\mu 
+ \sqrt{\frac{2}{3}} \; \bar\psi_n \; \partial_\mu \pi^0 \; 
\Psi_{\Delta 0}^\mu
+ \frac{1}{\sqrt{3}} \; \bar\psi_n \; 
\partial_\mu \pi^- \; \Psi_{\Delta +}^\mu
\right) \; + \; \mbox{h.c.}
\label{lag_delta}
\ee
The empirical value of the $\pi N\Delta$ coupling is $g_{\pi N \Delta} = 20.22$. 

The introduction of the $\Delta$ into chiral EFT could in principle be accompanied by ``new'' 
$\pi\pi NN$ contact terms in the Lagrangian, of the form of the second term in Eq.~(\ref{lag_n}),
or with other structures. Physically such terms describe short-distance degrees of freedom 
that are integrated out in the derivation of the effective theory. They contain dynamical 
information and cannot be determined from the symmetries alone. We do not introduce any 
explicit contact terms at this time; their contributions can easily be included later.
We shall justify this choice a posteriori, by showing that the large-$N_c$ relations between the
densities resulting from $N$ and $\Delta$ intermediate states are satisfied without explicit contact terms.
We note that this ambiguity is related to that resulting from the off-shell extension of
the propagator (see Sec.~\ref{subsec:spinors}). It will be seen that both result in identical 
contact term contributions to the transverse densities in the light-front representation. 
\section{Light-front representation with $\Delta$ isobar}
\label{sec:lightfront}
\subsection{Current matrix element}
\label{subsec:current}
We now derive the light-front representation of the peripheral transverse densities 
including the $\Delta$ isobar contribution, following the approach developed in Ref.~\cite{Granados:2015rra} 
for chiral EFT with nucleons only. The first step is to compute the peripheral contributions to the current 
matrix element Eq.~(\ref{me_general})
from the processes Fig.~\ref{fig:diag}a and b (nucleon) and Fig.~\ref{fig:diag}d (isobar) 
in terms of Feynman integrals, and to separate each one of them into two parts:
\begin{itemize}
\item[(i)] An ``intermediate-baryon'' term, which contains the intermediate-baryon pole of the triangle 
diagram at $l^2 = M_B^2$ ($B = N, \Delta$). This term will then be evaluated as a contour integral over the
light-front energy variable and represented as an overlap integral of $N \rightarrow \pi B$ 
light-front wave functions.
\item[(ii)] A ``contact'' term, which combines (a) contributions from the triangle graph in which the 
intermediate-baryon pole is canceled by factors $l^2 - M_B^2$ arising from the numerator of 
the Feynman integral; (b) contributions from the diagram involving the explicit $\pi\pi NN$ 
vertices in the Lagrangian.
\end{itemize}
The separation becomes unique if we require that the numerator of intermediate-baryon term do not 
contain finite powers in $l^2 - M_B^2$, i.e., that they are all absorbed in the contact term.
In performing the separation we use the specific off-shell behavior of the EFT propagators and vertices.
We also systematically neglect non-peripheral contributions to the current matrix element, i.e., 
terms that do not contribute to the two-pion cut.

The nucleon EFT contribution to the peripheral isovector current matrix element results from the two diagrams 
of Fig.~\ref{fig:diag}a and b and is given by
\be
\langle N_2 | \, J^\mu (0) \, | N_1 \rangle^V_N &\equiv & 
\langle N(p_2, \sigma_2) | \, J^\mu (0) \, | N(p_1, \sigma_1) \rangle^V_N
\nonumber
\\[2ex]
&=& \frac{g_A^2}{2 F_\pi^2} \; \int\!\frac{d^4 k}{(2\pi)^4} \; 
\; \frac{(-i) (2 k^\mu) \, \bar u_2 i \hat k_2 \gamma_5 (\hat l + M_N) i \hat k_1 \gamma_5 u_1}
{(k_1^2 - M_\pi^2 + i0) (k_2^2 - M_\pi^2 + i0) (l^2 - M_N^2 + i0)} 
\label{triangle_momentum_1}
\\[2ex]
&+& \frac{1}{2 F_\pi^2} \; \int \! \frac{d^4 k}{(2\pi )^4} 
\; \frac{ i (2 k^\mu) \, \bar u_2 \hat k u_1}{(k_1^2 - M_\pi^2 + i0)
(k_2^2 - M_\pi^2 + i0)} ,
\label{contact_momentum}
\ee
where $k_{1, 2} = k \mp \Delta, l = p_1 - k_1 = p_2 - k_2$, and the integration is over the
average pion 4-momentum $k$.
In the numerators the vector $2 k^\mu$ represents the vector current of the charged pion.
Using the anticommutation relations between the gamma matrices and the Dirac equation for 
the nucleon spinors the numerator of the triangle diagram, Eq.~(\ref{triangle_momentum_1}), 
can be rewritten as
\be
\frac{g_A^2}{2 F_\pi^2} \; \bar u_2 i \hat k_2 \gamma_5 (\hat l + M_N) i \hat k_1 \gamma_5 u_1
&=& \frac{2 g_A^2 M_N^2}{F_\pi^2} \; \bar u_2 \hat k u_1
\; + \; \frac{g_A^2}{2 F_\pi^2} (l^2 - M_N^2) [\bar u_2 \hat k u_1 \; + \; (...) ]
\label{bilinear_simplified}
\ee
In the second term the factor $(l^2 - M_N^2)$ cancels the nucleon pole in Eq.~(\ref{triangle_momentum_1}). 
This term thus results in an integral of the same form as the contact diagram, Eq.~(\ref{contact_momentum}).
The (...) in Eq.~(\ref{bilinear_simplified}) stands for terms that are even under 
$k \rightarrow -k$ and drop out in the integral, as the remaining integrand is odd under $k \rightarrow -k$,
cf.\ Eq.~(\ref{contact_momentum}). The total nucleon contribution can thus be represented as
\be
\langle N_2 | \, J^\mu (0) \, | N_1 \rangle^V_N 
&=& 
\int\!\frac{d^4 k}{(2\pi)^4} \; 
\; \frac{i (2 k^\mu) \, A_{N, \; {\rm interm}}}
{(k_1^2 - M_\pi^2 + i0) (k_2^2 - M_\pi^2 + i0) (l^2 - M_N^2 + i0)} 
\label{current_nucleon_interm}
\\[2ex]
&+& \int \! \frac{d^4 k}{(2\pi )^4} 
\; \frac{ i (2 k^\mu) \, A_{N, \; {\rm cont}}}{(k_1^2 - M_\pi^2 + i0)
(k_2^2 - M_\pi^2 + i0)} ,
\label{current_nucleon_contact}
\\[2ex]
A_{N, \; {\rm interm}} &=& - \frac{2 g_A^2 M_N^2}{F_\pi^2} \; \bar u_2 \hat k u_1 ,
\label{numerator_nucleon_interm}
\\[2ex]
A_{N, \; {\rm cont}} &=& \frac{1 - g_A^2}{2 F_\pi^2} \; \bar u_2 \hat k u_1 .
\label{numerator_nucleon_contact}
\ee
By construction, for 4-momenta on the nucleon mass shell, $l^2 = M_N^2$, 
the numerator Eq.~(\ref{numerator_nucleon_interm}) 
coincides with the numerator of the original triangle diagram, Eq.~(\ref{triangle_momentum_1}), 
\beq
A_{N, \; {\rm interm}}\, |_{l^2 = M_N^2} \;\; = \;\; - \frac{g_A^2}{2 F_\pi^2} \; 
[\bar u_2 i \hat k_2 \gamma_5 (\hat l + M_N) i \hat k_1 \gamma_5 u_1]_{l^2 = M_N^2} .
\label{numerator_nucleon_onshell}
\eeq

The isobar contribution to the peripheral current matrix element resulting from 
diagram Fig.~\ref{fig:diag}d is given by
\be
\langle N_2 | J^\mu (0) | N_1 \rangle^V_{\Delta}
&=& \frac{g_{\pi N\Delta}^2}{3 M_N^2} 
\int\frac{d^4 k}{(2\pi )^4}
\frac{(-i) (2 k^\mu) \, [\bar u_2 \, k_{2\alpha} R^{\alpha\beta} (l) k_{1\beta}\, u_1]}
{(k_2^2 - M_\pi^2 + i0)
(k_1^2 - M_\pi^2 + i0) (l^2 - M_\Delta^2 + i0)} .
\label{triangle_momentum_delta}
\ee
The bilinear form in the in the numerator is an invariant function of the 4-vectors 
$p_1, p_2$ and $k$ (for given spin quantum numbers of the external nucleon spinors, $\sigma_1$ and $\sigma_2$). 
Using the specific form of $R^{\alpha\beta}(l)$, as given by Eqs.~(\ref{projector_expanded})
and (\ref{projector_factorized}) without the mass-shell condition, and employing gamma
matrix identities and the Dirac equation for the external nucleon spinors, the bilinear form
can be represented as
\be
\bar u_2 \, k_{2\alpha} R^{\alpha\beta} (l) k_{1\beta}\, u_1
&=& \bar u_2 ( \mathcal{F} + \mathcal{G} \, \hat k) u_1 .
\ee
Here $\mathcal{F}$ and $\mathcal{G}$ are scalar functions of four independent invariants
formed from the 4-vectors $p_1, p_2$ and $k$. We choose as independent invariants $k_1^2, k_2^2, l^2$ and $t$,
and write
\beq
\mathcal{F}, \mathcal{G} \;\; = \;\; \textrm{functions} (k_1^2, \, k_2^2, \, l^2, \, t).
\eeq
Because the functions arise from the contraction of finite-rank tensors, their dependence on
the invariants is polynomial, and we can expand them around the position of the poles of the 
propagators in Eq.~(\ref{triangle_momentum_delta}). In the expansion in $k_{1}^2 - M_\pi^2$ 
and $k_{2}^2 - M_\pi^2$ we keep only the zeroth order term, as any finite power of 
$k_{1, 2}^2 - M_\pi^2$ would cancel the pole of one of the pion propagators and thus
no longer contribute to the two-pion cut (such contributions would be topologically equivalent
to those of the diagrams of Fig.~\ref{fig:diag}c, which we also neglect). In the expansion in
$l^2 - M_\Delta^2$ we keep all terms, and write the result as the sum of the zeroth order 
term and a remainder summing up the powers $(l^2 - M_\Delta^2)^k$ with $k \geq 1$.
In this way the bilinear form becomes
\be
\bar u_2 \, k_{2\alpha} R^{\alpha\beta} (l) k_{1\beta}\, u_1
&=& \bar u_2 ( F + G \, \hat k) u_1 \; + \; 
(l^2 - M_\Delta^2 ) 
\bar u_2 ( F_{\rm cont} + G_{\rm cont} \, \hat k) u_1 
\nonumber
\\[2ex]
&+& \; \textrm{terms $\sim (k_{1, 2}^2 - M_\pi^2)$},
\label{R_delta_on_off}
\ee
where
\be
F &\equiv& \mathcal{F}(k_1^2 = M_\pi^2, k_2^2 = M_\pi^2, l^2 = M_\Delta^2, t) ,
\\[2ex]
F_{\rm cont} &\equiv& \frac{\mathcal{F}(k_1^2 = M_\pi^2, k_2^2 = M_\pi^2, l^2, t) 
- \mathcal{F}(\ldots, \; l^2 = M_\Delta^2, \; \ldots)}{l^2 - M_\Delta^2},
\ee
and similarly for $G$ and $G_{\rm cont}$ in terms of $\mathcal{G}$.
The explicit expressions are
\be
F &\equiv& \phantom{-} \left[ \frac{t}{2} - M_N^2 + 
\frac{(M_\Delta^2 + M_N^2 - M_\pi^2)^2}{4 M_\Delta^2} \right]
(M_N + M_\Delta) 
\nonumber \\[2ex] 
&& 
- \; \frac{1}{3} 
\left( M_N + \frac{M_\Delta^2 + M_N^2 - M_\pi^2}{2 M_\Delta}\right)^2
(M_N - M_\Delta ) ,
\\[3ex]
G &\equiv& -\left[ \frac{t}{2} - M_N^2 + 
\frac{(M_\Delta^2 + M_N^2 - M_\pi^2)^2}{4 M_\Delta^2} \right]
\nonumber \\[2ex] 
&& + \; \frac{1}{3} 
\left( M_N + \frac{M_\Delta^2 + M_N^2 - M_\pi^2}{2 M_\Delta}\right)^2 ,
\label{f1_cut}
\\[2ex]
F_{\rm cont} 
&\equiv & \phantom{-}
\frac{1}{3 M_\Delta^2} ( M_\Delta^2 - M_N^2 + M_N M_\Delta + M_\pi^2 ) (M_N + M_\Delta )
\nonumber
\\[2ex]
&& \; + \frac{(l^2 - M_\Delta^2 )}{6 \, M_\Delta^2}  M_N ,
\\[2ex]
G_{\rm cont} 
&\equiv & 
- \frac{1}{3 M_\Delta^2} ( M_\Delta^2 - M_N^2 + M_N M_\Delta + M_\pi^2 )
\nonumber
\\[2ex]
&& \; - \frac{(l^2 - M_\Delta^2 )}{6 \, M_\Delta^2} .
\ee
The isobar contribution can thus be represented as
\be
\langle N_2 | \, J^\mu (0) \, | N_1 \rangle^V_\Delta 
&=& 
\int\!\frac{d^4 k}{(2\pi)^4} \; 
\; \frac{i \, (2 k^\mu) \, A_{\Delta, \; {\rm interm}}}
{(k_1^2 - M_\pi^2 + i0) (k_2^2 - M_\pi^2 + i0) (l^2 - M_N^2 + i0)} 
\label{current_delta_interm}
\\[2ex]
&+& \int \! \frac{d^4 k}{(2\pi )^4} 
\; \frac{ i (2 k^\mu) \, A_{\Delta, \; {\rm cont}}}{(k_1^2 - M_\pi^2 + i0)
(k_2^2 - M_\pi^2 + i0)} ,
\label{current_delta_contact}
\\[2ex]
A_{\Delta, \; {\rm interm}} &=& - \frac{g_{\pi N\Delta}^2}{3 M_N^2} \;
\bar u_2 ( F + G \, \hat k) u_1 ,
\label{numerator_delta_interm}
\\[2ex]
A_{\Delta, \; {\rm cont}} &=& - \frac{g_{\pi N\Delta}^2}{3 M_N^2} 
\; \bar u_2 (F_{\rm cont} + G_{\rm cont} \hat k) u_1 ,
\label{numerator_delta_contact}
\ee
which is analogous to Eqs.~(\ref{current_nucleon_interm})-(\ref{numerator_nucleon_contact}) for the nucleon.
By construction, for 4-momenta on the $\Delta$ mass shell, $l^2 = M_\Delta$, the numerator 
Eq.~(\ref{numerator_delta_interm}) coincides with the numerator of the 
original triangle diagram, Eq.~(\ref{triangle_momentum_delta}), 
\be
A_{\Delta, \; {\rm interm}} \, |_{l^2 = M_\Delta^2} &=& 
- \frac{g_{\pi N\Delta}^2}{3 M_N^2} 
[\bar u_2 \, k_{2\alpha} R^{\alpha\beta} (l) k_{1\beta}\, u_1]_{l^2 = M_N^2} .
\label{numerator_delta_onshell}
\ee

Combining nucleon and isobar contributions, the total peripheral current in chiral EFT can
be represented as
\be
\langle N_2 | \, J^\mu (0) \, | N_1 \rangle^V &=&  
\sum_{B = N, \Delta} \langle N_2 | \, J^\mu (0) \, | N_1 \rangle^V_B 
\\[2ex]
\langle N_2 | \, J^\mu (0) \, | N_1 \rangle^V_B 
&=& 
\int\!\frac{d^4 k}{(2\pi)^4} \; 
\; \frac{i \, (2 k^\mu) \, A_{B, \; {\rm interm}}}
{(k_1^2 - M_\pi^2 + i0) (k_2^2 - M_\pi^2 + i0) (l^2 - M_B^2 + i0)} 
\label{current_interm}
\\[2ex]
&+& \int \! \frac{d^4 k}{(2\pi )^4} 
\; \frac{ i (2 k^\mu) \, A_{B, \; {\rm cont}}}{(k_1^2 - M_\pi^2 + i0)
(k_2^2 - M_\pi^2 + i0)} ,
\label{current_contact}
\ee
with numerators given by the specific expressions Eqs.~(\ref{numerator_nucleon_interm}), 
(\ref{numerator_nucleon_contact}), (\ref{numerator_delta_interm}) and (\ref{numerator_delta_contact}).
These expressions will be used subsequently to derive the light-front wave function overlap 
representation of the current matrix element.

The intermediate-baryon term of the peripheral current is independent of the off-shell behavior of 
the propagator and vertices, both in the $N$ and $\Delta$ case. Because the numerator is
evaluated at $k_{1, 2}^2 = M_\pi^2$ (two-pion cut, or peripheral distances) and $l^2 = M_B^2$
(definition of intermediate-baryon term), the intermediate-baryon term effectively corresponds to 
the ``triple cut'' triangle diagram in which all internal lines are on 
mass-shell.\footnote{Such ``completely cut'' diagrams describe absorptive parts in unphysical regions
of the external kinematic variables and are used in the dispersive representation of multiparticle 
amplitudes (Mandelstam representation) \cite{LLIV}.} It thus expresses the physical particle content of 
the effective theory. The off-shell behavior enters only in the effective contact terms. 
In particular, the addition of an explicit new $\pi\pi NN$ contact term in 
connection with the $\Delta$ [cf.\ diagram Fig.~\ref{fig:diag}e] 
would modify only the effective contact term in Eq.~(\ref{current_contact}).

Some comments are in order regarding the ultraviolet (UV) behavior of the Feynman integrals.
As they stand, the Feynman integrals Eqs.~(\ref{current_nucleon_interm})-(\ref{numerator_nucleon_contact}) 
and Eqs.~(\ref{current_delta_interm})-(\ref{numerator_delta_contact}) are UV divergent.
In the dispersive approach the peripheral densities are computed using only the imaginary part 
of the form factors on the two-pion cut, which is UV finite.
In the present calculation the Feynman integrals could be regularized in a Lorentz-invariant manner, 
e.g. by subtraction of the integrals at $t = 0$; such subtractions would correspond to a modification
of the transverse density by delta functions at $b = 0$ and do not affect the peripheral densities.
In the wave function overlap representation derived in the following section, 
the regularization happens ``automatically'' when we change to the coordinate representation and 
compute the densities at non-zero transverse distance.
We thus do not need to explicitly regularize the integrals at this stage.
\subsection{Wave function overlap representation}
\label{subsec:overlap}
%
%
\begin{figure}[t]
\begin{center}
\includegraphics[width=.7\textwidth]{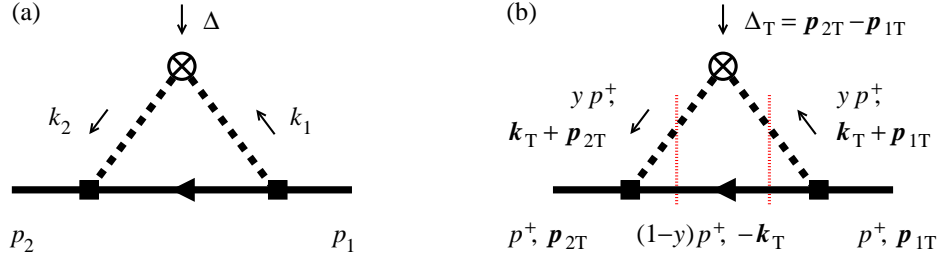}
\end{center}
\caption[]{Wave function overlap representation of the intermediate-baryon Feynman integral. 
(a) Feynman integral Eq.~(\ref{current_interm}) with 4-momenta. 
(b) Light-front components of momenta in frame where $\Delta$ is transverse, Eq.~(\ref{transverse_frame}).}
\label{fig:overlap}
\end{figure}
The next step is to represent the intermediate-baryon term of the current matrix element,
Eq.~(\ref{current_interm}), as an 
overlap integral of light-front wave functions describing the $N \rightarrow \pi B$ 
transition ($B = N, \Delta$) \cite{Granados:2015rra}. This is accomplished by performing a three-dimensional 
reduction of the Feynman integral using light-front momentum variables. The integral over the
light-front energy is performed by contour integration, closing the contour around the baryon pole, 
where the baryon momentum is on mass-shell and the numerator can be replaced by the 
on-shell projector.

The reduction is performed in a frame where the momentum transfer has only transverse components, 
cf.\ Eq.~(\ref{Delta_transverse}), such that (see Fig.~\ref{fig:overlap})
\be
p_1^+ \; = \; p_2^+ \; \equiv \; p^+, 
\hspace{2em} \bm{p}_{2T} - \bm{p}_{1T} \; = \; \bm{\Delta}_T,
\hspace{2em} p_{1}^- \; = \; \frac{M_N^2 + \bm{p}_{1T}^2}{p^+} ,
\hspace{2em} p_{2}^- \; = \; \frac{M_N^2 + \bm{p}_{2T}^2}{p^+} .
\label{transverse_frame}
\ee
Here $p^+ > 0$ is a free parameter which selects a particular frame in a class of 
frames related by longitudinal boosts. The transverse momenta satisfy 
$\bm{p}_{2T} - \bm{p}_{1T} = \bm{\Delta}_T$, while the overall transverse momentum 
remains unspecified. The loop momentum is described by its light-front components
$k^\pm \equiv k^0 \pm k^z$ and $\bm{k}_T \equiv (k^x, k^y)$,
and $k^+$ is parametrized in terms of the boost-invariant light-front momentum 
fraction of the pion,
\be
y \; \equiv \; k^+ / p^+ .
\ee
The integrand of the Feynman integral Eq.~(\ref{current_interm}) has simple poles 
in $k^-$ and can be computed by closing the contour around the pole of the baryon 
propagator \cite{Granados:2015rra}. Here it is essential that we have organized the 
contributions in Eqs.~(\ref{current_interm}) and (\ref{current_contact}) 
such that the integrand of Eqs.~(\ref{current_interm}) does not contain any non-zero 
powers of $l^2 - M_B^2$, which would be ``invisible'' at the baryon pole;
[they are contained in the contact term Eq.~(\ref{current_contact})].
At the baryon pole in $k^-$ the pion virtualities are, up to a factor, given by the 
invariant mass differences between the external nucleon and the pion-baryon
system in the intermediate state (here $\bar y \equiv 1 - y$),
\beq
- \frac{k_{1}^2 - M_\pi^2}{y} \;\; = \;\;  
\frac{(\bm{k}_T + \bar y \bm{p}_{1T})^2 + M_\pi^2}{y} + 
\frac{(\bm{k}_T + \bar y\bm{p}_{1T})^2 + M_B^2}{\bar y} - M_N^2 .
\;\; \equiv \;\; \Delta\mathcal{M}^2_{N \rightarrow \pi B} (y, \bm{k}_T, \bm{p}_{1T}) ,
\label{invariant_mass}
\eeq
and similarly for $k_1 \rightarrow k_2$ and $\bm{p}_{1T} \rightarrow \bm{p}_{2T}$.
This allows us to convert the pion denominators in the Feynman integral into light-front energy 
denominators for the $N \rightarrow \pi B$ transition and back.

Furthermore, at the baryon pole in $k^-$ the numerator of the Feynman integral
can be factorized. At the pole the baryon 4-momentum is on mass shell, $l^2 = M_N^2$. 
By construction the numerator of the intermediate-baryon Feynman integral
Eq.~(\ref{current_interm}) is equal to the numerator of the original Feynman integral 
on the baryon mass shell. At the same time, on the mass shell the residue of the
baryon propagator in the original Feynman integral can be replaced by the projector
on on-shell baryon spinors (or vector-spinors) with 4-momentum $l$.
For the intermediate nucleon contribution, using Eqs.~(\ref{numerator_nucleon_onshell})
and (\ref{projector_nucleon}), we get 
[here $u_1 \equiv u(p_1, \sigma_1), u_2 \equiv u(p_2, \sigma_2)$ and $u \equiv u(l, \sigma)$]
\be
A_{N, \; {\rm interm}} 
&=& - \frac{g_A^2}{2 F_\pi^2} \; 
[\bar u_2 i \hat k_2 \gamma_5 (\hat l + M_N) i \hat k_1 \gamma_5 u_1 ]_{l^2 = M_N^2}
\\[2ex]
&=& - \frac{g_A^2}{2 F_\pi^2} \; \sum_\sigma \;
[\bar u_2 i \hat k_2 \gamma_5 u] [\bar u i \hat k_1 \gamma_5 u_1] 
\hspace{2em}
({\textstyle \sigma = \pm\frac{1}{2}}) 
\label{numerator_nucleon_1}
\\[2ex]
&=& 2 \, \sum_\sigma \; \Gamma^\ast_{\pi NN} (y, \bm{k}_T, \bm{p}_{2T}; \sigma, \sigma_2)
\; \Gamma_{\pi NN} (y, \bm{k}_T, \bm{p}_{1T}; \sigma, \sigma_1)
\hspace{2em}
({\textstyle \sigma = \pm\frac{1}{2}}) .
\ee
In the last step we have expressed the bilinear forms in terms of the general $\pi NN$
vertex function, which is regarded here as a function of the light-front momentum variables 
characterizing the on-shell nucleon 4-momenta. It is defined as
\beq
\Gamma_{\pi NN} (y, \bm{k}_T, \bm{p}_{1T}; \sigma, \sigma_1) \;\; \equiv \;\;
- \frac{g_A}{2 F_\pi} \; \bar u(l, \sigma) i \hat k_1 \gamma_5 u (p_1, \sigma_1) ,
\label{vertex_nucleon}
\eeq
where the light-front components of the nucleon 4-momenta $p_1$ and $l$ are 
given by [cf.~Eq.~(\ref{transverse_frame})]
\be
p_1^+ &=& p^+, \;\;
\bm{p}_{1T}, \;\;
p_1^- = \frac{|\bm{p}_{1T}|^2 + M_N^2}{p^+} ,
\label{vertex_p1}
\\[1ex]
l^+ &=& \bar y p^+, \;\;
\bm{l}_T = - \bm{k}_T , \;\; 
l^- = \frac{|\bm{k}_T|^2 + M_N^2}{\bar y p^+} , 
\label{vertex_l}
\ee
and the pion 4-momentum $k_1$ is given by the difference of the nucleon 4-momenta, 
$k_1 = p_1 - l$, or
\be
k_1^+ &=& y p^+, \;\; \bm{k}_{1T} = \bm{p}_{1T} - \bm{l}_T, \;\; k_1^- = p_1^- - l^-.
\label{vertex_k1}
\ee
We emphasize that the assignment of the pion 4-momentum follows unambiguously from our 
reduction procedure. The second bilinear form in Eq.~(\ref{numerator_nucleon_1}) is 
directly given by Eq.~(\ref{vertex_nucleon}); the first bilinear form is given by the 
complex conjugate of the same function, evaluated at arguments 
$\bm{p}_{1T} \rightarrow \bm{p}_{2T}$ and $\sigma_1 \rightarrow \sigma_2$.
Because both nucleon 4-momenta at the $\pi NN$ vertices
are on mass shell, the vertices can equivalently be written as pseudoscalar vertices, cf.\
Eq.~(\ref{axial_pseudoscalar}).

For the intermediate isobar contribution, using Eqs.~(\ref{numerator_delta_onshell})
and (\ref{projector_delta}), we get
\be
A_{\Delta, \; {\rm interm}} &=& - \frac{g_{\pi N\Delta}^2}{3 M_N^2} 
 [\bar u_2 \, k_{2\alpha} R^{\alpha\beta} (l) k_{1\beta}\, 
u_1]_{l^2 = M_\Delta^2}
\\[1ex]
&=& - \frac{g_{\pi N\Delta}^2}{3 M_N^2} \; \sum_\sigma \;
 [\bar u_2 \, k_{2\alpha}  U^\alpha (l)] [\bar U^\beta(l) k_{1\beta}\, u_1]
\\[2ex]
&=& - \frac{2}{3} 
\sum_\sigma \; \Gamma^\ast_{\pi N\Delta} (y, \bm{k}_T, \bm{p}_{2T}; \sigma, \sigma_2)
\; \Gamma_{\pi N\Delta} (y, \bm{k}_T, \bm{p}_{1T}; \sigma, \sigma_1)
\hspace{2em}
({\textstyle \sigma = \pm\frac{3}{2}, \pm\frac{1}{2}}) .
\ee
In the last step we have introduced the general vertex function for the 
$\pi N \Delta$ transition, defined as
\beq
\Gamma_{\pi N\Delta} (y, \bm{k}_T, \bm{p}_{1T}; \sigma, \sigma_1) \;\; \equiv \;\;
\frac{g_{\pi N\Delta}}{\sqrt{2} M_N} \;
\bar U^\beta (l, \sigma) k_{1\beta} u (p_1, \sigma_1) ,
\label{vertex_delta}
\eeq
where the light-front components of the nucleon 4-momenta $p_1$ and $l$ are given by the same 
expressions as Eqs.~(\ref{vertex_p1}) and (\ref{vertex_l}), the only difference being
that the minus component of the isobar momentum is
\be
l^- = \frac{|\bm{k}_T|^2 + M_\Delta^2}{\bar y p^+} . 
\ee
The pion 4-momentum $k_1$ is again given by the difference of the baryon 4-momenta, 
Eq.~(\ref{vertex_k1}).

In summary, the intermediate-baryon part of the peripheral current matrix element,
Eq.~(\ref{current_interm}), can be expressed as
\be
\langle N_2 | \, J^+ (0) \, | N_1 \rangle^V_{B, \; {\rm interm}}
&=& \frac{p^+}{2\pi} \int_0^1\frac{dy}{y\bar y} \int \frac{d^2 k_T}{(2\pi)^2}
\nonumber \\[2ex]
&\times& C_B \;
\sum_{\sigma} \Psi^\ast_{N \rightarrow \pi B} (y, \bm{k}_T, \bm{p}_{2T}; \sigma, \sigma_2) \;
\Psi_{N \rightarrow \pi B} (y, \bm{k}_T, \bm{p}_{1T}; \sigma, \sigma_1) 
\hspace{2em}
(B = N, \Delta),
\label{me_interm_overlap}
\ee
where
\beq
C_N \;\; = \;\; 2, \hspace{2em} C_\Delta \;\; = \;\; -2/3
\label{isospin_factors}
\eeq
are isospin factors, and 
\be
\Psi_{N \rightarrow \pi B} (y, \bm{k}_T, \bm{p}_{1T}; \sigma, \sigma_{1}) 
&\equiv& 
\frac{\Gamma_{\pi N B} (y, \bm{k}_T, \bm{p}_{1T}; \sigma, \sigma_{1})}
{\Delta\mathcal{M}^2_{N \rightarrow \pi B}(y, \bm{k}_T, \bm{p}_{1T})} 
\label{psi_def}
\ee
is the light-front wave function of the $N \rightarrow \pi B$ transition, consisting
of the vertex function and the invariant-mass denominator.
Equation~(\ref{me_interm_overlap}) represents the Feynman integral as an overlap integral 
of the light-front wave functions describing the transition from the initial nucleon 
state $N_1$ to the $\pi B$ intermediate state, and back to the final nucleon state $N_2$.

Some comments are in order regarding the isospin structure. In our convention the $\pi N B$ 
vertex and the light-front wave function are normalized such that they describe the transition between 
the highest-isospin states of the initial nucleon and the intermediate baryon multiplets,
which are
\beq
\begin{array}{ll}
\textrm{intermediate $N$:} \hspace{1em} & p \;\; \rightarrow \;\; p \; + \; \pi^0 ,
\\[1ex]
\textrm{intermediate $\Delta$:} & p \;\; \rightarrow \;\; \Delta^{++} \; + \; \pi^- . 
\end{array}
\eeq
The factors $C_B$ account for the total contribution of configurations with an
intermediate baryon and pion to the nucleon's isovector current, including the probability 
for the configuration (i.e., the square of the coupling constant) and the pion charge. 
To determine the values, let us assume that the external nucleon state is the proton.
In the case of intermediate $N$, the only configuration contributing to the current is 
$n + \pi^+$ (with relative coupling $\sqrt{2}$), 
and the isospin factor is $C_N = \sqrt{2} \times \sqrt{2} = 2$.
In the case of intermediate $\Delta$, the possible configurations are
$\Delta^{++} + \pi^-$ (with relative coupling 1)
and $\Delta^0 + \pi^+$ (with relative coupling $1/\sqrt{3}$), and the isospin factor is 
$C_\Delta = (-1 + \frac{1}{3})/(-1) = \frac{2}{3}$.

\begin{figure}[t]
\begin{center}
\includegraphics[width=.75\textwidth]{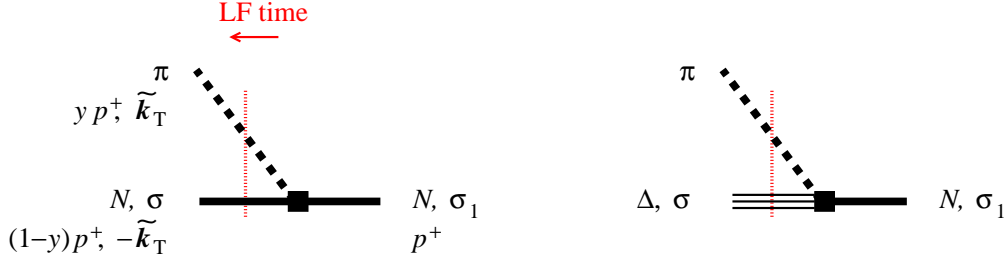}
\end{center}
\caption[]{Light-front wave functions of $N \rightarrow \pi B$ transition in chiral EFT 
($B = N, \Delta$). The spin quantum number and the light-front momenta of the particles 
are indicated. The transverse momenta correspond to the transverse rest frame, 
Eq.~(\ref{psi_restframe}).}
\label{fig:diag_wf}
\end{figure}

The light-front wave functions of the nucleons moving with transverse momenta 
$\bm{p}_{1T}$ and $\bm{p}_{2T}$ can be expressed in terms of those in the transverse 
rest frame \cite{Granados:2015rra}. The relation between the wave functions follows
from the Lorentz invariance of the vertex function and the invariant-mass denominator
and takes the form
\be
\Psi_{N \rightarrow \pi B} (y, \bm{k}_T, \bm{p}_{1T}; \sigma, \sigma_1) &=& 
\Psi (y, \bm{k}_T + \bar y \bm{p}_{1T}, \, \bm{0}; \sigma, \sigma_1) 
\;\; \equiv \;\;
\Psi (y, \bm{k}_T + \bar y \bm{p}_{1T}; \sigma, \sigma_1) .
\ee
For simplicity we denote the light-front wave function in the transverse rest frame by the 
same symbol as that of the moving nucleon, only dropping the overall transverse 
momentum argument. The explicit expressions for the rest frame wave function can be 
obtained by setting $\bm{p}_{1T} = 0$ in the invariant
mass difference Eq.~(\ref{invariant_mass}) and the vertices,
Eqs.~(\ref{vertex_nucleon}) and (\ref{vertex_delta}). They are:
\be
\Psi_{N \rightarrow \pi B} (y, \widetilde{\bm{k}}_T; \sigma, \sigma_1)
&\equiv& 
\frac{\Gamma_{\pi N B} (y, \widetilde{\bm{k}}_T; \sigma, \sigma_1)}
{\Delta\mathcal{M}^2_{N \rightarrow \pi B} (y, \widetilde{\bm{k}}_T)} 
\hspace{2em} (B = N, \Delta),
\label{psi_restframe}
\\[1ex]
\Delta\mathcal{M}^2_{N \rightarrow \pi B} (y, \widetilde{\bm{k}}_T) &\equiv&
\frac{\widetilde{\bm{k}}_T^2 + M_\pi^2}{y} + 
\frac{\widetilde{\bm{k}}_T^2 + M_B^2}{\bar y} - M_N^2 ,
\label{invariant_mass_restframe}
\\[1ex]
\Gamma_{\pi NN} (y, \widetilde{\bm{k}}_T; \sigma, \sigma_1) &\equiv&
- \frac{g_A}{2 F_\pi} \; \bar u(l, \sigma) i (\hat p_1 - \hat l) \gamma_5 u (p_1, \sigma_1) ,
\label{vertex_nucleon_restframe}
\\[1ex]
\Gamma_{\pi N\Delta} (y, \widetilde{\bm{k}}_T; \sigma, \sigma_1) & \equiv &
\frac{g_{\pi N\Delta}}{\sqrt{2} M_N} \;
\bar U^\mu (l, \sigma) (p_1 - l)_\mu u (p_1, \sigma_1) ,
\label{vertex_delta_restframe}
\\[1ex]
p_1^+ &=& p^+, \;\; \bm{p}_{1T} = 0, \;\; p_1^- = \frac{M_N^2}{p^+} ,
\label{vertex_p1_restframe}
\\[1ex]
l^+ &=& \bar y p^+, \;\; \bm{l}_T = - \widetilde{\bm{k}}_T , \;\; 
l^- = \frac{|\widetilde{\bm{k}}_T|^2 + M_B^2}{\bar y p^+} , 
\label{vertex_l_restframe}
\ee
where $\widetilde{\bm{k}}_T$ denotes the transverse momentum argument of the rest frame
wave function (see Fig.~\ref{fig:diag_wf}). In terms of the rest-frame wave functions
the peripheral current matrix element is then given by
\be
\langle N_2 | \, J^+ (0) \, | N_1 \rangle^V_{B, \; {\rm interm}}
&=& 
\frac{p^+}{2\pi} \int_0^1\frac{dy}{y\bar y} \int \frac{d^2 k_T}{(2\pi)^2}
\nonumber \\
&\times& \; C_B \; \sum_{\sigma} 
\Psi^\ast_{N \rightarrow \pi B} (y, \bm{k}_T + \bar y \bm{p}_{2T}; \sigma , \sigma_2) \;
\Psi_{N \rightarrow \pi B} (y, \bm{k}_T + \bar y \bm{p}_{1T}; \sigma, \sigma_1) 
\hspace{2em} (B = N, \Delta) . \hspace{2em}
\label{me_interm_overlap_restframe}
\ee
%
%
\subsection{Polarization states}
\label{subsec:polarization}
To evaluate the wave function overlap formulas we need to specify the polarization states 
for the nucleon and isobar spinors. It is natural to describe the polarization of
the particles using light-front helicity states \cite{Brodsky:1997de}. In this formulation one
constructs the spinor/vector for a particle with light-front momentum $l^+$ and $\bm{l}_T$
by starting with the spinor/vector in the particle rest frame ($l^+ = M_B, \bm{l}_T = 0$),
then performing a longitudinal boost to the desired longitudinal momentum $l^+ \neq M_B$, 
then a transverse boost to $\bm{l}_T \neq 0$. The spinors/vectors thus defined correspond to definite
polarization states in the particle rest frame and transform in a simple way under longitudinal 
and transverse boosts.

The bispinors for a spin-1/2 particle of mass $M_B$ with definite light-front helicity 
$\sigma = \pm 1/2$ are given explicitly by \cite{Brodsky:1997de,Leutwyler:1977vy}
\beq
u(l, \sigma) \;\; = \;\; 
\frac{1}{\sqrt{2 l^+}} \left[ l^+ \gamma^- 
+ (M_B - \bm{\gamma}_T \cdot \bm{l}_{T} ) \gamma^+ \right] 
\left( \begin{array}{c} \chi (\sigma ) \\[1ex] 0 \end{array} \right) ,
\label{spinor_boosted}
\eeq
where $\chi (\sigma)$ are the rest-frame 2-spinors for polarization in the positive 
and negative $z$-direction,
\beq
\chi(\sigma = 1/2) \; = \; 
\left( \begin{array}{c} 1 \\ 0 \end{array}\right),
\hspace{2em}
\chi(\sigma = -1/2) \; = \; 
\left( \begin{array}{c} 0 \\ 1 \end{array}\right) .
\label{chi_along_z}
\eeq
The bispinors Eq.~(\ref{spinor_boosted}) satisfy the normalization and completeness 
relations, Eqs.~(\ref{normalization_nucleon}) and (\ref{projector_nucleon}) 
for the nucleon ($B = N$), or Eqs.~(\ref{spinor_delta_normalization}) and (\ref{spinor_delta_projector})
for the isobar ($B = \Delta$). Similarly, the polarization vectors for a spin-1 particle with 
definite light-front helicity $\lambda = (1, 0, -1)$ are given by
(see e.g.\ Ref.~\cite{Dziembowski:1997vh})
\be
\epsilon^+(\lambda, l) &=& \frac{l^+}{M_\Delta} \alpha^z (\lambda) ,
\nonumber
\\[1ex]
\epsilon^-(\lambda, l) &=& \frac{l_T^2-M_\Delta^2}{M_\Delta l^+} \alpha^z (\lambda)
\; + \; \frac{2 \bm{l}_T}{l_+} \cdot \bm{\alpha}_T(\lambda) ,
\nonumber 
\\[1ex]
\bm{\epsilon}_T (\lambda, l) &=& \frac{\bm{l}_T}{M_\Delta}
\alpha^z (\lambda) \; + \; \bm{\alpha}_T (\lambda) ,
\label{polvec}
\ee
where $\bm{\alpha}(\lambda) \equiv (\alpha^z(\lambda), \bm{\alpha}_T(\lambda))$ 
denotes the polarization 3-vector of the particle in the rest frame,
\beq
\epsilon (\lambda, l)  \;\; = \;\; (0, \bm{\alpha}(\lambda))
\hspace{2em} [\textrm{rest frame}, \; l = (M_\Delta, 0)] ,
\label{rfpvec}
\eeq
and describes states with circular polarization along the $z$-axis,
\beq
\bm{\alpha}(\lambda = \pm 1) \; = \; \frac{1}{\sqrt{2}}
\left( \begin{array}{c} \mp 1 \\ -i \\ 0 \end{array}\right),
\hspace{2em}
\bm{\alpha}(\lambda = 0) \; = \; 
\left( \begin{array}{c} 0 \\ 0 \\ 1 \end{array}\right) .
\label{alpha_along_z}
\eeq
The vector-bispinors for a spin-3/2 particle with definite light-front helicity are then 
defined by Eq.~(\ref{vector_bispinor}). The coupling of the spinors to spin-3/2 
is performed in rest frame, and the bispinor and vector are boosted independently
to the desired light-front momentum.

Using the light-front helicity spinors we can now compute the components of the 
$\pi N B$ vertex function in the nucleon rest frame. The results
are conveniently expressed as bilinear forms between the rest-frame spinors/vectors
characterizing the states. For the transition to a nucleon ($B = N$) we obtain 
from Eq.~(\ref{vertex_nucleon_restframe})
\be
\Gamma_{\pi N N} (y, \widetilde{\bm{k}}_T; \sigma, \sigma_1)
&=& \frac{2i g_A M_N}{F_\pi \sqrt{\bar{y}}}
\left[ y M_N \, S^z (\sigma, \sigma_1) + 
\widetilde{\bm{k}}_T \cdot \bm{S}_T(\sigma, \sigma_1) 
\right] ,
\label{vertex_nucleon_longit}
\ee
where $S^z$ and $\bm{S}_T \equiv (S^x, S^y)$ are the components of the 3-vector 
characterizing the spin transition matrix element in the rest frame
\be
S^i (\sigma, \sigma_1) &\equiv& \chi^\dagger (\sigma)
({\textstyle\frac{1}{2}} \sigma^i ) \chi (\sigma_1) 
\hspace{2em} (i = x, y, z),
\label{S_vector}
\ee
in which $\sigma^i$ are the Pauli matrices. The first term in Eq.~(\ref{vertex_nucleon_longit})
is diagonal in the nucleon light-front helicity, $S^z (\sigma, \sigma_{1}) = \sigma \, 
\delta(\sigma, \sigma_{1})$, and describes a transition 
to a $\pi N$ state with orbital angular momentum projection $L^z = 0$. 
The second term in Eq.~(\ref{vertex_nucleon_longit}) is off-diagonal in light-front
helicity and describes a transition to a state in which the pion has orbital angular 
momentum projection $L^z = \pm 1$.
For the transition to the isobar ($B = \Delta$) the vertex given by 
Eq.~(\ref{vertex_delta_restframe}) can be written as
\beq
\Gamma_{\pi N \Delta} (y, \widetilde{\bm{k}}_T; \sigma, \sigma_1)
\;\; = \;\; \sum_{\lambda\sigma'} C^{3/2, \sigma}_{1, \lambda;\; 1/2, \sigma'} \;
\bm{\alpha}(\lambda) \cdot \bm{\mathcal{P}} (y, \widetilde{\bm{k}}_T, \sigma', \sigma_1) ,
\label{vertex_delta_restframe_1}
\eeq
where the 3-vector $\bm{\mathcal{P}} \equiv (\mathcal{P}^z, \bm{\mathcal{P}}_T)$ 
represents the bilinear form in the bispinors in the vertex function. 
Its components are obtained as
\be
\left.
\begin{array}{cr}
\mathcal{P}^z
\\[3ex]
\bm{\mathcal{P}}_T
\end{array}
\right\} (y, \widetilde{\bm{k}}_T, \sigma', \sigma_1) \;
&=& \; \frac{\sqrt{2} g_{\pi N\Delta}}{M_N \bar y^{3/2}}
\left\{ \bar{M} \; \delta (\sigma', \sigma_1) + i \widetilde{\bm{k}}_T \cdot 
[ \bm{e}_z \times \bm{S}_T(\sigma', \sigma_1) ] \right\}
\; \left\{
\begin{array}{cr} 
\displaystyle
\frac{\widetilde{\bm{k}}_T^2}{2 M_\Delta} - M_d
\\[2ex]
- \widetilde{\bm{k}}_T \end{array}
\right\} ,
\label{vertex_delta_restframe_2}
\ee
in which 
\be
\bar{M} &\equiv& \frac{M_\Delta+\bar{y}M_N}{2},
\\[1ex]
M_d &\equiv& \frac{M_\Delta^2-\bar{y}^2M^2_N}{2M_\Delta}.
\label{MtlD}
\ee
\subsection{Coordinate-space wave function}
\label{subsec:coordinate}
The peripheral transverse densities are conveniently expressed in terms of the
transverse coordinate-space wave functions of the $N \rightarrow \pi B$ transition. 
They are defined as the transverse Fourier transform of the momentum-space wave 
functions at fixed longitudinal momentum fraction $y$,
\be
\Phi_{N \rightarrow \pi B} (y, \bm{r}_T, \sigma, \sigma_1) 
&\equiv&
\int \frac{d^2\widetilde{k}_{T}}{(2\pi)^2} \; 
e^{i \bm{\widetilde{k}}_T \cdot \bm{r}_T} \; 
\Psi_{N \rightarrow \pi B} (y, \bm{\widetilde{k}}_T; \sigma, \sigma_1) 
\label{psi_coordinate}
\\[1ex]
&=&
\int \frac{d^2\widetilde{k}_{T}}{(2\pi)^2} \; 
e^{i \bm{\widetilde{k}}_T \cdot \bm{r}_T} \; 
\frac{\Gamma_{\pi N B} (y, \widetilde{\bm{k}}_T; \sigma, \sigma_{1})}
{\Delta\mathcal{M}^2_{N \rightarrow \pi B}(y, \widetilde{\bm{k}}_T)} .
\label{psi_coordinate_explicit}
\ee
The vector $\bm{r}_T$ represents the relative transverse coordinate, i.e., the difference 
of the transverse positions of the $\pi$ and $B$, which are regarded as point particles
in the context of chiral EFT. The wave function thus describes the physical transverse 
size distribution of the $\pi B$ system in the intermediate state. The invariant-mass 
denominator in Eq.~(\ref{psi_coordinate_explicit}) is given by 
Eq.~(\ref{invariant_mass_restframe}) and can be written in the form
\be
\Delta\mathcal{M}^2_{N \rightarrow \pi B}(y, \widetilde{\bm{k}}_T)
 &=& \frac{\bm{\widetilde{k}}_T^2 + M_{T, B}^2}{y \bar y} ,
\label{invariant_mass_transverse}
\ee
where $M_{T, B} \equiv M_{T, B} (y)$ 
is the $y$-dependent transverse mass governing the transverse 
momentum dependence,
\be
M_{T, B}^2 &\equiv& \bar{y} M_\pi^2 \; +\; y (M_B^2 - \bar{y} M_N^2) 
\nonumber
\\[2ex]
&=& \bar{y} M_\pi^2 \; + \; y^2 M_N^2 \; + \; y (M_B^2 - M_N^2) .
\label{M_T_def}
\ee
The Fourier transform Eq.~(\ref{psi_coordinate_explicit}) can be expressed in terms 
of modified Bessel functions using the identity
\beq
\int \frac{d^2\widetilde{k}_{T}}{(2\pi)^2} \; 
\; \frac{e^{i \bm{r}_T \cdot \bm{\widetilde{k}}_T}}
{\widetilde{\bm{k}}_T^2 + M_{T, B}^2} \;\; = \;\; \frac{K_0 (M_{T, B}\, r_T)}{2\pi}
\hspace{3em} (r_T \equiv |\bm{r}_T|),
\eeq
from which formulas with additional powers of $\widetilde{\bm{k}}_T$ in the numerator
can be derived by differentiating with respect to the vector $\bm{r}_T$. 
The coordinate-space wave functions can be represented as sums of
spin structures depending on the transverse unit vector 
$\bm{n}_T \equiv \bm{r}_T/r_T$, multiplied by radial functions
depending on the modulus $r_T$.

For the nucleon intermediate state ($B = N$) the coordinate-space wave 
function is of the form \cite{Granados:2015rra}
\be
\Phi_{N \rightarrow \pi N} (y,\bm{r}_T, \sigma, \sigma_1) \; &=& \;
-2i S^z(\sigma, \sigma_1) \; U_0(y, r_T) \; + \; 2 \, \bm{n}_T
\cdot \bm{S}_T (\sigma, \sigma_1) \; U_1(y, r_T) 
\nonumber
\\[1ex]
&& (\bm{n}_T \equiv \bm{r}_T / r_T) .
\label{coordinate_nucleon}
\ee
The two terms in Eq.~(\ref{coordinate_nucleon}) represent
structures with definite orbital angular momentum around the $z$-axis 
in the rest frame, namely $L^z = 0$ and $L^z = \pm 1$. The radial wave 
functions are obtained as
\be
\left.
\begin{array}{r}
U_0(y,r_T) 
\\[2ex]
U_1(y,r_T)
\end{array}
\right\}
&=& \frac{g_A M_N \, y \sqrt{\bar{y}}}{2\pi F_\pi}
\left\{
\begin{array}{r}
y M_N \; K_0(M_{T, N} r_T)
\\[2ex]
M_{T, N} \; K_1(M_{T, N} r_T)
\end{array}
\right\} .
\label{radial_nucleon}
\ee
For the isobar intermediate state ($B = \Delta$) the coordinate-space
wave function is of the form
\beq
\Phi_{N \rightarrow \pi \Delta} (y,\bm{r}_T, \sigma, \sigma_1) \;\; = \;\; 
\sum_{\lambda\sigma'} C^{3/2, \sigma}_{1, \lambda;\; 1/2, \sigma'} \;
\bm{\alpha}(\lambda) \cdot \bm{\mathcal{R}} (y, \bm{r}_T, \sigma', \sigma_1) ,
\label{coordinate_delta}
\eeq
where $\bm{\mathcal{R}} \equiv (\mathcal{R}^z, \bm{\mathcal{R}}_T)$ is a 
3-vector-valued function, the components of which are given by
\be
\mathcal{R}^z (y, \bm{r}_T, \sigma', \sigma_1)
&=& 
- V_0(y,r_T) \; \delta
\; + \; 2 \, \bm{n}_T \cdot (\bm{e}_z \times \bm{S}_T) V_1(y,r_T) ,
\label{R_z}
\\[2ex]
\bm{\mathcal{R}}_T (y, \bm{r}_T, \sigma', \sigma_1)
&=& - \; 2 i (\bm{e}_z\times \bm{S}_T) \, W_0 (y,r_T)
- i \bm{n}_T \, W_1 (y,r_T) \; \delta
\; - \; 2 i \, 
(\bm{n}_T \otimes \bm{n}_T - {\textstyle\frac{1}{2}} \bm{1}) 
\cdot (\bm{e}_z\times \bm{S}_T) 
\, W_2 (y,r_T) 
\nonumber 
\\[2ex]
&& [\bm{n}_T \equiv \bm{r}_T / r_T, \; 
\delta \equiv \delta(\sigma', \sigma_1), \;
\bm{S}_T \equiv \bm{S}_T(\sigma', \sigma_1)] .
\label{R_T}
\ee
$\mathcal{R}^z$ contains structures with orbital angular momentum $L^z = 0$ 
and $\pm 1$, whereas $\bm{\mathcal{R}}_T$ contains structures with 
$L^z = 0, \pm 1$ and $\pm 2$. The radial functions are now obtained as
\be
\left.
\begin{array}{r}
V_0 \\[3.7ex]
V_1 \\[3.7ex]
W_0 \\[3.7ex]
W_1 \\[3.7ex]
W_2 
\end{array}
\right\}
(y, r_T)
&=& \frac{g_{\pi N\Delta} y}{2 \pi \sqrt{2} M_N \sqrt{\bar y}}
\left\{\begin{array}{r} 
\displaystyle
2 \bar{M} \left( \frac{M_{T\Delta}^2}{2M_\Delta}+M_d \right)
K_0(M_{T\Delta} r_T)
\\[2.5ex]
\displaystyle
M_{T\Delta} \left( \frac{M_{T\Delta}^2}{2M_\Delta} + M_d \right)
K_1(M_{T\Delta} r_T)
\\[3.5ex]
\frac{1}{2} M_{T\Delta}^2 \;  K_0 (M_{T\Delta} r_T)
\\[3.5ex]
2 \bar{M} M_{T\Delta} 
K_1(M_{T\Delta} r_T)
\\[3.5ex]
M_{T\Delta}^2 K_2(M_{T\Delta} r_T)
\end{array} \; \right \} . 
\label{radial_delta}
\ee

The coordinate-space wave functions fall off exponentially at large 
transverse distances $r_T$, with a width that is determined by the
inverse transverse mass Eq.~(\ref{M_T_def}) and depends on the pion momentum 
fraction $y$. At large values of the argument the modified Bessel 
functions behave as
\beq
K_{n} (M_{T, B} r_T) \;\; \sim \;\; 
\sqrt{\frac{\pi}{2}} \; \frac{e^{-M_{T, B} r_T}}{\sqrt{M_{T, B} r_T}}
\hspace{2em} (M_{T, B} r_T \gg 1; \; n = 0, 1, 2, \ldots) .
\label{K01_asymptotic}
\eeq
This behavior is caused by the singularity of the momentum-space wave function 
at (complex) transverse momenta $\bm{\widetilde{k}}_T^2 = - M_{T, B}^2$, 
cf.~Eq.~(\ref{invariant_mass_transverse}), which corresponds to the vanishing of
the invariant mass denominator of the wave function, $\Delta \mathcal{M}^2 = 0$.

\begin{figure}
\begin{tabular}{ll}
\includegraphics[width=.48\textwidth]{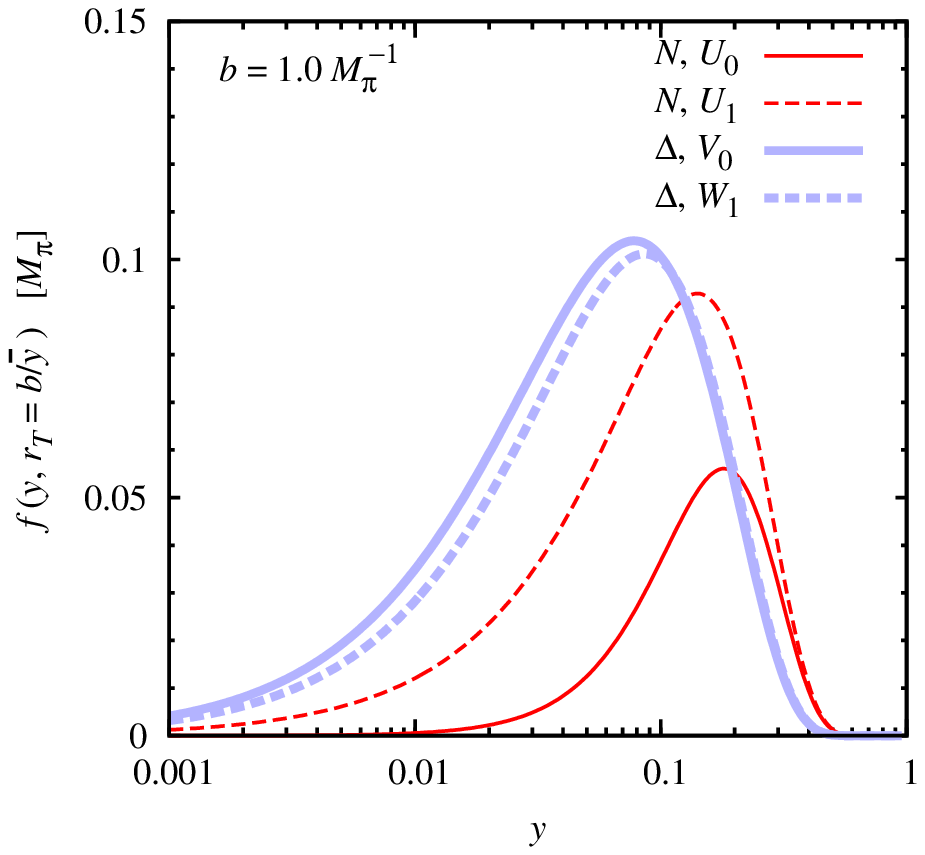}
&
\includegraphics[width=.48\textwidth]{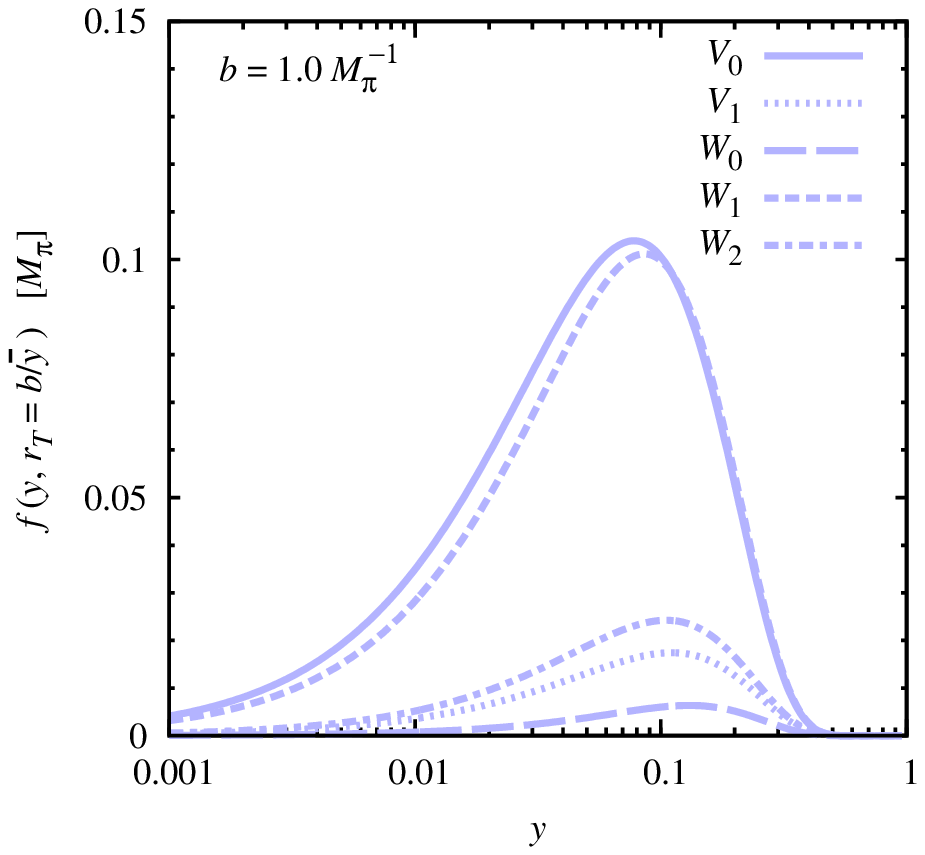}
\\[-2ex]
(a) & (b)
\end{tabular}
\caption[]{Chiral light-front wave functions in coordinate representation. 
(a) Radial wave functions of $\pi N$ system ($U_0, U_1$) and $\pi \Delta$ system 
(dominant $V_0, W_1$), as functions of the pion momentum fraction $y$, at transverse 
separation $r_T = b/\bar y$ with $b = 1.0 \, M_\pi^{-1}$.
(b) Full set of radial wave functions of $\pi \Delta$ system.}
\label{fig:wfy}
\end{figure}

The chiral $N \rightarrow \pi B$ light-front wave functions are defined in the 
parametric domain
\beq
y \;\; = \;\; O(M_\pi / M_N),
\hspace{2em} 
|\bm{r}_T| \;\; = \;\; O(M_\pi^{-1}) ,
\hspace{2em} \textrm{or} \hspace{2em}
|\bm{\widetilde{k}}_T| \;\; = \;\; O(M_\pi) ,
\label{parametric_chiral}
\eeq
where they describe peripheral pions carrying a parametrically small fraction of
the nucleon's longitudinal momentum, and are to be used in this domain only.
In the nucleon rest frame this corresponds to the domain in which all components of 
the pion 4-momentum are $O(M_\pi)$ (``soft pion'') and chiral dynamics is valid.
In the case of the nucleon intermediate state ($B = N$), Eq.~(\ref{M_T_def}) shows
that for $y = O(M_\pi / M_N)$ the transverse mass is
\beq
M_{T, N} \;\; = \;\; O(M_\pi) 
\hspace{2em} \textrm{for} \hspace{2em}
y = O(M_\pi/M_N),
\label{transverse_mass_nucleon_parametric}
\eeq
so that the exponential range of the coordinate-space wave function is
$O(M_\pi^{-1})$, cf.~Eq.~(\ref{K01_asymptotic}). In the case of the 
isobar intermediate state ($B = \Delta$) the situation is more complex, as a
an additional scale is present in the $N$-$\Delta$ mass splitting. The transverse 
mass squared Eq.~(\ref{M_T_def}) now contains a term 
\beq
y (M_\Delta^2 - M_N^2) \;\; = \;\; y (M_\Delta - M_N) (M_\Delta + M_N) ,
\eeq
which is of ``mixed'' order $O[M_\pi (M_\Delta - M_N)]$ for regular chiral
momentum fractions $y = O(M_\pi / M_N)$, and becomes $O(M_\pi^2)$
only for parametrically small fractions $y = O\{M_\pi^2/[(M_\Delta - M_N) M_N]\}$.
In the strict chiral limit $M_\pi \rightarrow 0$ at fixed $M_\Delta - N_N$ 
the isobar intermediate state would therefore be a short-distance contribution 
to the densities, or parametrically suppressed at distances $O(M_\pi^{-1})$.
However, at the physical value of $M_\pi$ the $O[M_\pi (M_\Delta - M_N)]$ term
in Eq.~(\ref{M_T_def}) is numerically of the same magnitude as the $O(M_\pi^2)$ ones,
and the intermediate isobar contribution is comparable to the intermediate nucleon one, 
as is seen in the numerical results below (see Fig.~\ref{fig:wfy}a).

For pion longitudinal momentum fractions $y = O(1)$ the transverse mass Eq.~(\ref{M_T_def})
is $M_T = O(M_B)$ for both $B = N$ and $\Delta$ intermediate states, and the
transverse range of the wave functions is $O(M_B^{-1})$. This region does not
correspond to chiral dynamics, and the wave functions defined by the above
expressions have no physical meaning there. In the calculation of the
peripheral transverse densities this region of $y$ does not contribute,
as the wave functions will be evaluated at distances $r_T = b/(1 - y)$
and therefore vanish exponentially in the limit $y \rightarrow 1$, 
cf.~Eq.~(\ref{K01_asymptotic}). In the calculations
in Sec.~\ref{sec:densities} we can thus formally extend the integral up to $y = 1$
without violating the parametric restriction Eq.~(\ref{parametric_chiral}).
This self-regulating property is a major advantage of our coordinate-space
formulation. The selection of large transverse distances $b = O(M_\pi^{-1})$
automatically enforces the parametric restrictions in the wave function overlap 
integral, and no explicit regulators are required.\footnote{The intermediate
isobar wave functions Eq.~(\ref{radial_delta}) contain a prefactor $1/\sqrt{\bar y}$
and become infinite when taking the limit $y \rightarrow 1$ {\it at fixed separation} $r_T$.
This singularity is purely formal and occurs outside of the parametric region 
of applicability. In the calculation of the transverse densities the wave functions 
are evaluated {\it at moving separation} $r_T = b/\bar y$ and vanish exponentially
in the limit $y \rightarrow 1$, cf.~Eq.~(\ref{K01_asymptotic}) and
Figs.~\ref{fig:wfy}a and b.} 

%
%
Numerical results for the $N \rightarrow \pi B$ wave functions are shown in 
Figs.~\ref{fig:wfy}a and b. Figure~\ref{fig:wfy}a shows the radial functions
for the intermediate nucleon, $U_0$ and $U_1$, and the two dominant radial functions
for the isobar, $V_0$ and $W_1$, as functions of the pion longitudinal momentum fraction $y$.
The functions are evaluated at a moving transverse separation $r_T = b/\bar y$,
in the same way as they enter in the calculation of the transverse densities 
(see Sec.~\ref{subsec:peripheral_densities}). Several features are worth
noting: (a)~The nucleon radial functions are centered around  values $y \sim M_\pi/ M_N$.
The isobar wave functions are shifted toward slightly smaller $y$, due to
the different behavior of the transverse mass. (b)~The radial functions at 
fixed $b$ vanish exponentially for $y \rightarrow 1$. (c)~The helicity-conserving
($L^z = 0$) nucleon function $U_0$ and the helicity-flip ($L^z = 1$) function $U_1$
are of the same magnitude at non-exceptional $y$. In the limit $y \rightarrow 1$ 
the two functions approach each other; in the limit $y \rightarrow 0$ the helicity-conserving 
function vanishes more rapidly than the helicity-flip one. The same pattern is
observed for the helicity-conserving and -flip isobar wave functions, $V_0$ and $W_1$.
(d)~Intermediate nucleon and isobar wave functions are generally of the same magnitude.

Figure~\ref{fig:wfy}a shows the full set of radial wave functions for the
intermediate isobar, Eq.~(\ref{radial_delta}). It is seen that the functions
$V_0$ and $W_1$ are numerically dominant over $V_1, W_0$ and $W_2$. This pattern 
follows from the fact that $\bar M \gg M_{T, \Delta}$ at non-exceptional values 
of $y$ and can be explained more formally in the large-$N_c$ limit 
(see Sec.~\ref{sec:largenc}).
\subsection{Peripheral densities}
\label{subsec:peripheral_densities}
The final step now is to calculate the peripheral transverse densities in terms of the
coordinate-space light-front wave functions \cite{Granados:2015rra}. To this end we 
consider the general form factor decomposition of current matrix element Eq.~(\ref{me_general}) 
in the frame where momentum transfer $\Delta$ is transverse, Eq.~(\ref{transverse_frame}), 
and with the nucleon spin states chosen as light-front helicity states 
Eq.~(\ref{spinor_boosted}),
\be
\langle N_2 | \, J^+ (0) \, | N_1 \rangle
&\equiv& 
\langle N(p^+, \bm{p}_{2T}, \sigma_2 ) |
\, J^+ (0) \, | N(p^+, \bm{p}_{1T}, \sigma_1 ) \rangle
\nonumber
\\[1ex]
&=& 
2p^+ \left[ \delta(\sigma_2, \sigma_1) \; F_1 (-\bm{\Delta}_T^2) 
\; - \; 
i(\bm{e}_z \times \bm{\Delta}_T)\cdot
\bm{S}_T (\sigma_2, \sigma_1) \; \frac{F_2 (-\bm{\Delta}_T^2)}{M_N}
\right] .
\label{me_plus}
\ee
The spin vector $\bm{S}_T (\sigma_2, \sigma_1)$ here is defined as in Eq.~(\ref{S_vector}),
but with the rest-frame 2-spinors describing the initial and final nucleon.
The term containing the form factor $F_1$ is diagonal in the nucleon light-front 
helicity, while the term with $F_2$ is off-diagonal. Explicit expressions for the 
form factors $F_1$ and $F_2$ are thus obtained by taking the diagonal and off-diagonal 
light-front helicity components, or, more conveniently, by multiplying Eq.~(\ref{me_plus}) with
$\delta(\sigma_1, \sigma_2)$ and 
$(\bm{e}_z \times \bm{\Delta}_T)\cdot \bm{S}_T (\sigma_1, \sigma_2)$ and summing over
$\sigma_1$ and $\sigma_2$,
\be
\left. 
\begin{array}{l}
F_1 (-\bm{\Delta}_T^2)  \\[4ex] F_2 (-\bm{\Delta}_T^2) 
\end{array}
\right\}
&=& \frac{1}{2p^+} \; \sum_{\sigma_1\sigma_2}
\langle N_2 | \, J^+ (0) \, | N_1 \rangle
\left\{\begin{array}{c}
\frac{1}{2} \delta(\sigma_1, \sigma_2) 
\\[2ex]
\displaystyle 
\frac{2 M_N}{\bm{\Delta}_T^2} \;
i (\bm{e}_z \times \bm{\Delta}_T)
\cdot \bm{S}_T(\sigma_1, \sigma_2)
\end{array}
\right\} .
\label{FF_matrix}
\ee
The peripheral current matrix element is expressed as an overlap integral of the
momentum-space $N \rightarrow \pi B$ light-front wave functions in 
Eq.~(\ref{me_interm_overlap_restframe}). In terms of the coordinate-space wave 
functions it becomes ($B = N, \Delta$)
\be
\langle N_2 | \, J^+ (0) \, | N_1 \rangle^V_{B, \; {\rm interm}}
&=& 
\frac{p^+}{2\pi} \int_0^1\frac{dy}{y\bar y} \int d^2 r_T \;
e^{i\bar y \bm{r}_T \bm{\Delta}_T}
\; C_B \; \sum_{\sigma} 
\Phi_{N \rightarrow \pi B}^\ast (y, \bm{r}_T; \sigma , \sigma_2) \;
\Phi_{N \rightarrow \pi B} (y, \bm{r}_T; \sigma, \sigma_1) .
\label{me_overlap_restframe_coordinate}
\ee
Note that the momentum transfer $\bm{\Delta}_T$ is Fourier-conjugate to $\bar y \bm{r}_T$;
this follows from the particular shift in the arguments of the rest-frame momentum-space 
wave functions and is a general feature of light-front kinematics. 
Substituting Eq.~(\ref{me_overlap_restframe_coordinate})
in Eq.~(\ref{FF_matrix}) and calculating the Fourier transform according to Eq.~(\ref{rho_def}), 
we obtain the intermediate-baryon contribution
to the transverse densities\footnote{For ease of notation we omit the label 
``interm'' on the intermediate-baryon contribution
to the densities and agree that $\rho_{1, B}^V(b) \equiv \rho_{1, B}^V(b)_{\rm interm}$
in the following. Explicit labels ``interm'' and ``cont'' will be used in 
Sec.~\ref{sec:largenc}, where we include the contact term contribution.
\label{foot:interm}}
 ($B = N, \Delta$)
\be
\left. 
\begin{array}{r}
\rho_{1, B}^V(b) \\[3ex]
(\bm{e}_z \times \bm{n}_T)  \; \widetilde\rho_{2, B}^V(b) 
\end{array}
\right\}
&=& 
\frac{C_B}{4\pi} \int_0^1\frac{dy}{y\bar y^3} \; \sum_{\sigma_1,\sigma_2,\sigma}
\displaystyle \Phi_{N \rightarrow \pi B}^\ast (y, \bm{r}_T; \sigma, \sigma_2) \;
\Phi_{N \rightarrow \pi B} (y, \bm{r}_T; \sigma, \sigma_1)
\;
\left\{
\begin{array}{c}
\frac{1}{2} \delta(\sigma_1, \sigma_2) 
\\[3ex]
\bm{S}_T (\sigma_1, \sigma_2)
\end{array}
\right\} 
\label{rho_overlap_phi}
\\[1ex]
&& (\bm{n}_T = \bm{b}/|\bm{b}|, \; \bm{r}_T = \bm{b}/\bar y).
\nonumber
\ee
In the representation Eq.~(\ref{rho_overlap_phi}) 
transverse rotational invariance is not manifest; however, the
densities are in fact rotationally invariant, as the angular factors are compensated 
by the angular dependence of wave function. Specifically, if we project the equation
for the spin-dependent density on the $y$-direction,
it becomes
\be
\cos\phi  \; \widetilde\rho_2^V(b) 
\; &=& \;
\frac{C_B}{4\pi} \int_0^1\frac{dy}{y\bar y^3} \; \sum_{\sigma_1,\sigma_2,\sigma}
\displaystyle \Phi_{N \rightarrow \pi B}^\ast (y, \bm{r}_T; \sigma, \sigma_2) \; 
\Phi_{N \rightarrow \pi B} (y, \bm{r}_T; \sigma, \sigma_1) \;
S^y (\sigma_1, \sigma_2)
\nonumber
\\[1ex]
&& (\bm{r}_T = \bm{b}/\bar y),
\label{rho_overlap_phi_alt}
\ee
in accordance with Eq.~(\ref{j_plus_rho}) and the visualization 
in Fig.~\ref{fig:densities}a. Note that the expressions in Eqs.~(\ref{rho_overlap_phi}) 
and (\ref{rho_overlap_phi_alt}) are diagonal in the pion longitudinal momentum 
fraction $y$ and the transverse separation $\bm{r}_T$, and thus represent 
true densities of the light-front wave functions of the intermediate $\pi B$ systems
\cite{Burkardt:2000za,Miller:2007uy,Miller:2010nz}.

We can now express the transverse densities in terms of the radial wave functions,
by substituting in Eqs.~(\ref{rho_overlap_phi}) and (\ref{rho_overlap_phi_alt})
the explicit form of the coordinate-space wave functions, Eqs.~(\ref{coordinate_nucleon})
and (\ref{coordinate_delta}), and performing the spin sums. The intermediate nucleon 
and isobar contributions are obtained as
\be
\left. 
\begin{array}{l}
\rho_{1, N}^V(b) \\[3ex] \widetilde\rho_{2, N}^V(b) 
\end{array}
\right\}
&=& 
\frac{C_N}{4\pi} \int_0^1\frac{dy}{y\bar y^3}
\left\{
\begin{array}{c}
\displaystyle U_0^2 \, + \, U_1^2
\\[3ex]
\displaystyle
- 2 \, U_0 \, U_1
\end{array}
\right\} ,
\label{rho_overlap_n}
\\[2ex]
\left.
 \begin{array}{cr}
 \rho_{1, \Delta}^V(b)
\\[3ex]
 \widetilde{\rho}_{2, \Delta}^V(b) 
 \end{array}\right\}&=& \frac{C_\Delta}{6\pi}
\int \frac{dy}{y\bar{y}^3}
 \left\{\begin{array}{cr}
  (V_0 + W_0)^2 + (V_1 + {\textstyle\frac{1}{2}} W_1)^2 
+ \frac{3}{4} (W_1^2 + W_2^2)
\\[3ex]
2 (V_0 + W_0) (V_1 + {\textstyle\frac{1}{2}}W_1) + \frac{3}{2} W_1 W_2
 \end{array}
 \right\} 
 \label{rho_overlap_delta}
\\[2ex]
&& \left[ U_0 \;\; \equiv \;\; U_0 (y, r_T = b/\bar y), \; \textrm{etc.} \right] .
\nonumber
\ee
Two properties of these expressions are worth noting: (a) Rotational invariance 
of the densities is now manifest, as the angular-dependent factors have canceled
in the calculation. (b) The spin-independent density $\rho_{1, B}^V$ involves
products of functions with the same orbital angular momentum, while the spin-dependent 
density $\widetilde\rho_{2, B}^V$ involves products of functions differing by
one unit of angular momentum; i.e., one has the selection rules ($B = N, \Delta$)
\beq
\left.
\begin{array}{ll}
\rho_{1, B}^V: \hspace{2em} & \Delta L^z = 0 \\[1ex]
\widetilde \rho_{2, B}^V: & \Delta L^z = 1
\end{array}
\right\} .
\eeq
This appears natural in view of the light-front helicity structure of the form factors, Eq.(\ref{FF_matrix}).

%
%
\begin{figure}
\begin{tabular}{l}
\includegraphics[width=.48\textwidth]{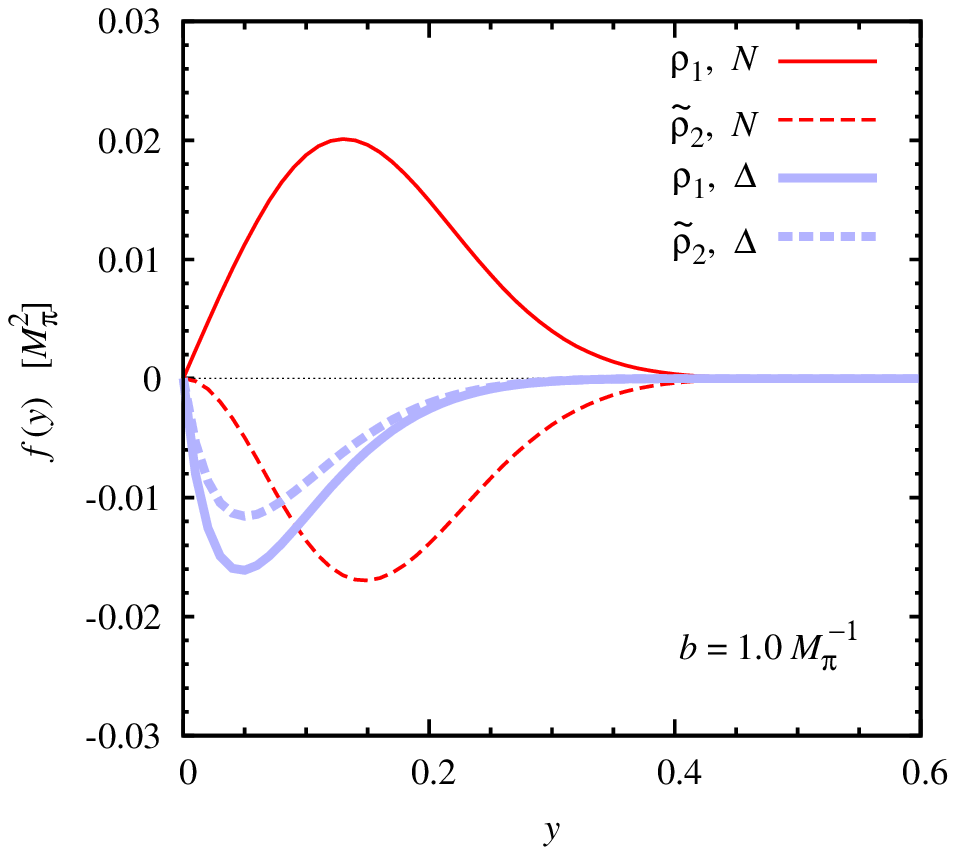}
\\[-4ex]
{\bf (a)}
\\[2ex]
\includegraphics[width=.48\textwidth]{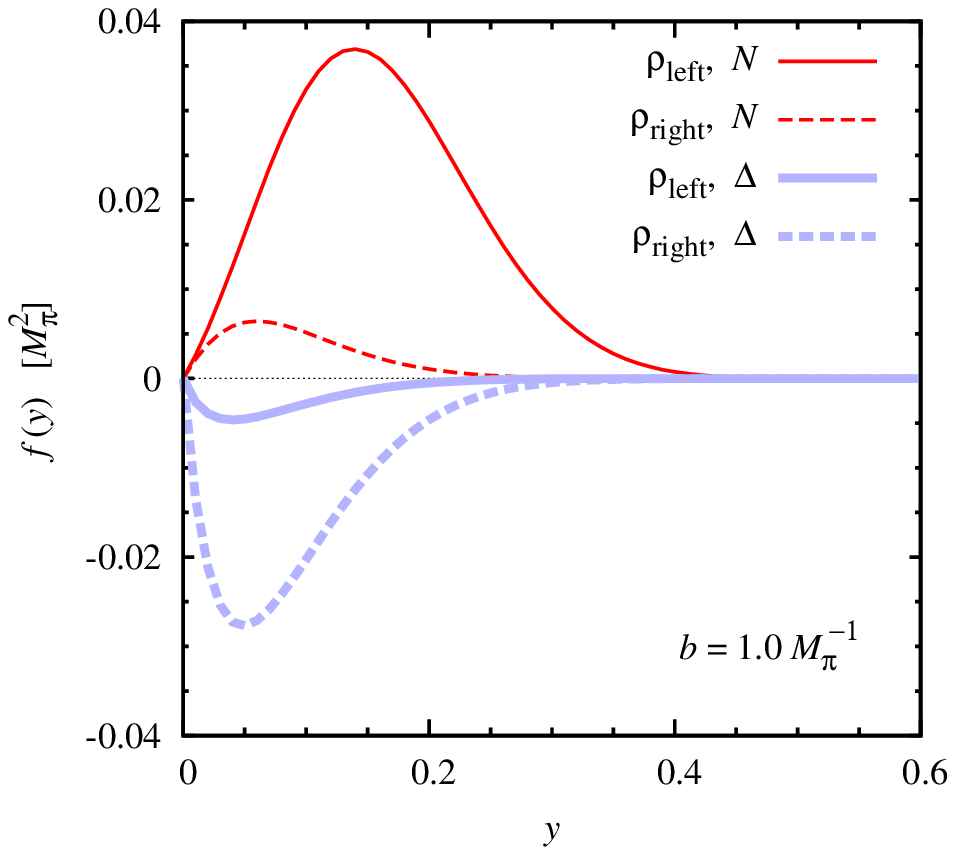}
\\[-4ex]
{\bf (b)}
\end{tabular}
\caption[]{Integrands of the light-front wave function overlap integrals for the
peripheral transverse densities, as functions of the pion momentum fraction $y$, 
at a distance $b = 1.0 \, M_\pi^{-1}$. 
(a) Integrands of the $N$ and $\Delta$ contributions to the spin-independent 
and -dependent densities, $\rho_1^V$ and $\widetilde\rho_2^V$, 
Eqs.~(\ref{rho_overlap_n}) and (\ref{rho_overlap_delta}). (b) Same for the left and 
right densities, $\rho_{\rm left}^V$ and $\rho_{\rm right}^V$, 
Eqs.~(\ref{rho_left_right_longitudinal}) and (\ref{rho_left_right_longitudinal_delta}).
The plots show the integrands including the 
factor $1/(y\bar y^3)$ and the spin/isospin-dependent prefactors, such that they give 
an impression of the actual distribution of strength in the integrals.}
\label{fig:densy}
\end{figure}
%
%
\begin{figure}
\begin{tabular}{l}
\includegraphics[width=.48\textwidth]{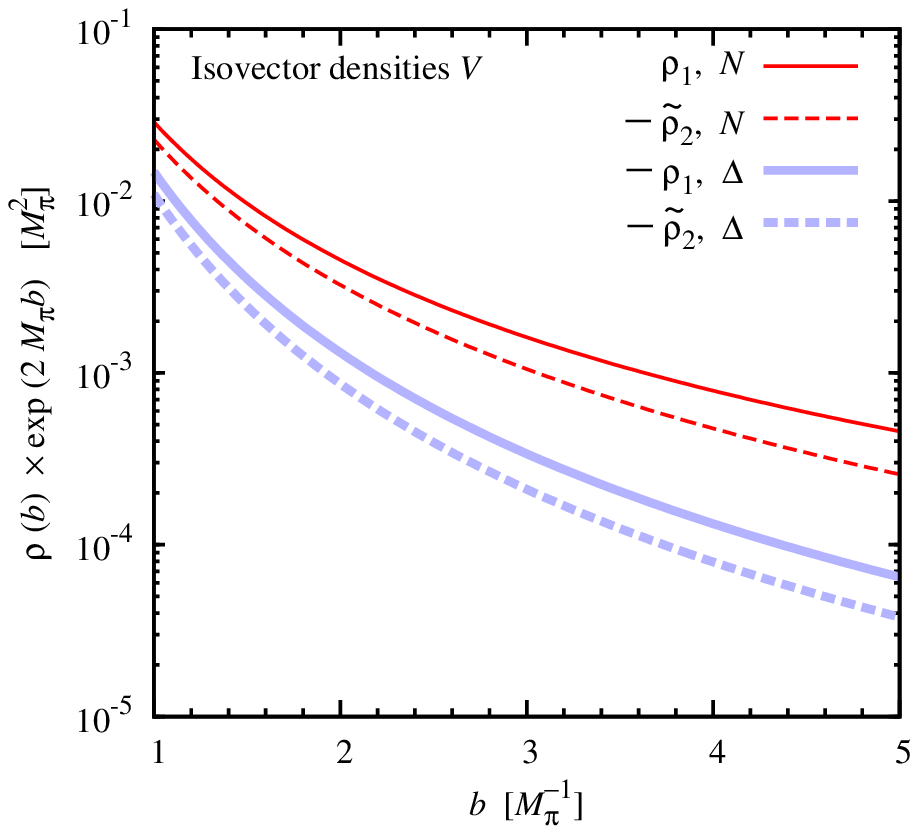}
\\[-4ex]
{\bf (a)}
\\[2ex]
\includegraphics[width=.48\textwidth]{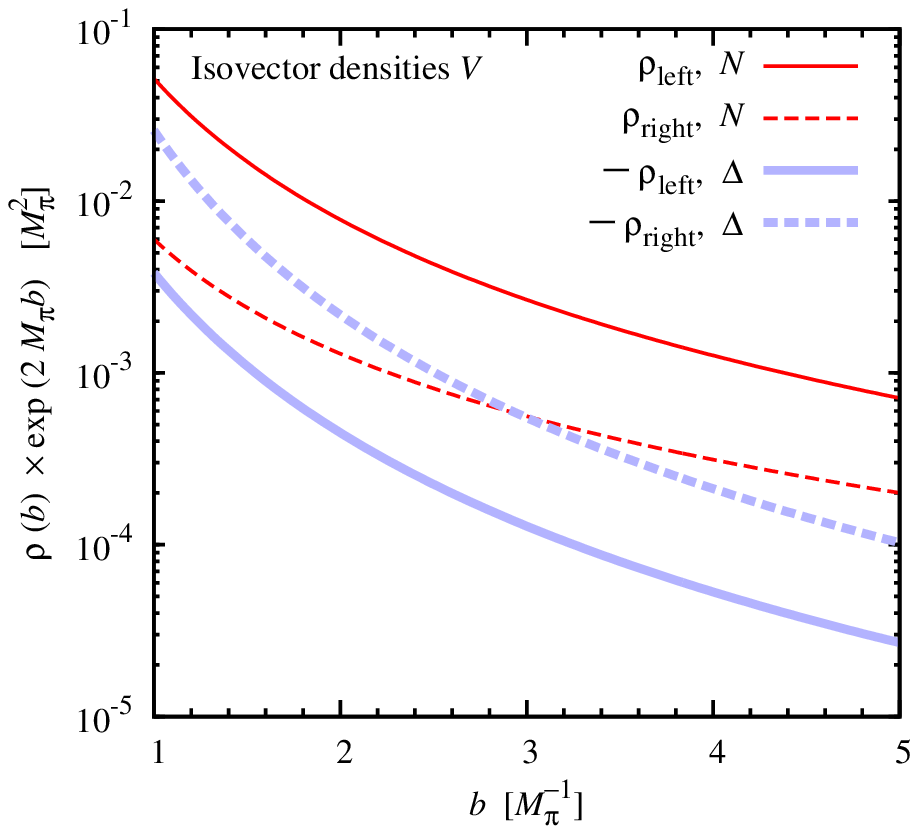}
\\[-4ex]
{\bf (b)}
\end{tabular}
\caption[]{Peripheral transverse densities obtained from the light-front wave function overlap integrals.
The plots show the densities rescaled by $\exp (2 M_\pi b)$, as functions of the distance $b$, 
in units of the pion mass $M_\pi$. (a) Intermediate $N$ and $\Delta$ contributions to the 
spin-independent and -dependent densities, $\rho_1^V$ and $\widetilde\rho_2^V$,
Eqs.~(\ref{rho_overlap_n}) and (\ref{rho_overlap_delta}). 
All contributions are plotted as positive functions in order to combine them in the same
graph (see the minus signs in the legend). (b) Same for the left and right densities, $\rho_{\rm left}^V$ 
and $\rho_{\rm right}^V$, Eqs.~(\ref{rho_left_right_longitudinal}) 
and (\ref{rho_left_right_longitudinal_delta}).}
\label{fig:densb}
\end{figure}

The left and right current densities, Eq.~(\ref{rho_left_right}), are readily obtained
as the sum and difference of the spin-independent and -dependent densities
in Eqs.~(\ref{rho_overlap_n}) and (\ref{rho_overlap_delta}),
\be
\rho^V_{{\rm left/right}, N}(b) \;\; = \;\; \rho_{1, N}^V \mp \widetilde{\rho}_{2, N}^V
\; &=& \; \frac{C_N}{4\pi}
\int \frac{dy}{y\bar{y}^3} \; (U_0 \pm U_1)^2 ,
\label{rho_left_right_longitudinal}
\\[2ex]
\rho^V_{{\rm left/right}, \Delta}(b) \;\; = \;\; 
\rho_{1, \Delta}^V \mp \widetilde{\rho}_{2, \Delta}^V
\; &=& \; \frac{C_\Delta}{6\pi}
\int \frac{dy}{y\bar{y}^3} \; \left[ (V_0 + W_0 \mp 
V_1 \mp {\textstyle\frac{1}{2}} W_1)^2
\; + \; {\textstyle\frac{3}{4}} (W_1 \mp W_2)^2 \right]
\label{rho_left_right_longitudinal_delta}
\\[3ex]
&& \left[ U_0 \;\; \equiv \;\; U_0 (y, r_T = b/\bar y), \; \textrm{etc.} \right] .
\nonumber
\ee
The integrands in both expressions are explicitly positive. Taking into account the
signs of the isospin factors, Eq.~(\ref{isospin_factors}), we conclude that
\be
\rho^V_{{\rm left/right}, N}(b) \; &>& \; 0 , 
\label{positivity_n}
\\[2ex]
\rho^V_{{\rm left/right}, \Delta}(b) \; &<& \; 0 .
\label{positivity_delta}
\ee
This generalizes the positivity condition for the intermediate-nucleon contribution
derived in Ref.~\cite{Granados:2015rra}. The definiteness conditions 
Eqs.~(\ref{positivity_n}) and (\ref{positivity_delta}) represent new insights
into the structure of the transverse densities gained from  the light-front representation, 
as they rely essentially on the expression of the densities as quadratic forms in 
the wave function. The conditions are essential for the
quantum-mechanical interpretation of the transverse densities as the current carried by a
quasi-free free peripheral pion (see Sec.~\ref{sec:mechanical}).

We can now evaluate the densities numerically using the overlap formulas.
Figure~\ref{fig:densy}a shows the integrand of the spin-independent and -dependent
densities as a function of $y$, at a typical chiral distance $b = 1.0 \, M_\pi^{-1}$.
One observes that: (a) The intermediate nucleon contributes positively to $\rho_1^V$
and negatively to $\widetilde{\rho}_2^V$, while the intermediate isobar contributes negatively
to both $\rho_1^V$ and $\widetilde{\rho}_2^V$. The $N$ and $\Delta$ contributions thus
tend to cancel in $\rho_1^V$, while they add in $\widetilde{\rho}_2^V$.
(b) The spin-dependent and -independent
densities satisfy $|\widetilde{\rho}_{2, B}^V| < |\rho_{1, B}^V|$,
and the absolute values are very close $|\widetilde{\rho}_{2, B}^V| \approx |\rho_{1, B}^V|$,
for both $B = N, \Delta$. 

Figure~\ref{fig:densy}b shows the integrand of the corresponding left and right transverse 
densities as a function of $y$. The nucleon and isobar contributions obey the definiteness
conditions Eqs.~(\ref{positivity_n}) and (\ref{positivity_delta}). One also observes 
significant differences in absolute value between the left and right densities, with
$|\rho_{{\rm left}, N}| \gg |\rho_{{\rm right}, N}|$ for the intermediate nucleon, and
$|\rho_{{\rm left}, \Delta}| \ll |\rho_{{\rm right}, \Delta}|$ for the isobar.
This strong left-right asymmetry can naturally be explained in the quantum-mechanical
picture in the rest frame (see Sec.~\ref{sec:mechanical}) and is a consequence of
the relativistic motion of pions in chiral 
dynamics \cite{Granados:2015lxa,Granados:2015rra}.

Figures~\ref{fig:densb}a and b show the densities themselves as functions of $b$.
The densities are plotted as positive functions (note the legend) and with the 
exponential factor $\exp (-2 M_\pi b)$ removed, i.e., the functions shown 
correspond to the pre-exponential factor in Eq.~(\ref{large_b_general}). 
One sees that (a) the densities generated by the same intermediate state 
($B = N, \Delta$) have approximately the same $b$-dependence. 
(b) The isobar contribution decreases faster at large $b$, as is expected on 
grounds of the larger mass of the intermediate state. (c) The left and right densities
in Fig.~\ref{fig:densb}b show the large left-right asymmetry observed
already in the integrand.

We emphasize that the densities obtained from the light-front wave function overlap integrals,
Eqs.~(\ref{rho_overlap_n}) and (\ref{rho_overlap_delta}),
are numerically identical to those obtained by evaluating the intermediate-baryon part 
of the original Feynman integrals, Eq.~(\ref{current_interm}), using the dispersive 
technique of Ref.~\cite{Granados:2013moa}. This represents a crucial test of the reduction
procedure of Sec.~\ref{subsec:current} and ensures the overall equivalence of the light-front
representation with the invariant chiral EFT results.
\section{Mechanical picture with $\Delta$ isobar}
\label{sec:mechanical}
\subsection{Transverse polarization}
A particularly simple representation of the transverse charge and magnetization densities 
is obtained by choosing transverse spin states for the external nucleons and the 
intermediate baryon. The densities $\rho_1^V$ and $\widetilde{\rho}_2^V$ are diagonal in 
transverse spin and permit a simple interpretation as the expectation value of the current
generated by a peripheral pion in a single transverse spin 
state \cite{Granados:2015lxa,Granados:2015rra}. We can now extend this picture to
include the isobar contributions to the densities.

Transversely polarized particle states in the light-front formulation
are constructed by preparing a transversely polarized state in the rest frame 
(we choose the $y$-direction)
and performing the longitudinal and transverse boosts to the desired light-front 
momentum (cf.~Sec.~\ref{subsec:polarization}). The bispinor for a spin-1/2 particle
with transverse polarization is obtained from the general formula
Eq.~(\ref{spinor_boosted}) by choosing the rest-frame 2-spinors as eigenspinors 
of the transverse spin operator $S^y = \frac{1}{2}\sigma^y$,
\beq
\chi_{\rm tr} (\tau = 1/2) \; = \; \frac{1}{\sqrt{2}}
\left( \begin{array}{c} 1 \\ i \end{array}\right),
\hspace{2em}
\chi_{\rm tr} (\tau = -1/2) \; = \; \frac{1}{\sqrt{2}}
\left( \begin{array}{c} i \\ 1 \end{array}\right) ,
\label{chi_along_y}
\eeq
where we use $\tau = \pm 1/2$ to denote the $y$-spin eigenvalues. The polarization vector 
for a spin-1 particle with transverse polarization is obtained
from Eqs.~(\ref{polvec}) and (\ref{rfpvec}) by choosing the rest-frame 3-vectors
as eigenvectors of the transverse angular momentum operator $L^y$, whose matrix representation is 
\beq
(L^y)^{jk} = (-i) \epsilon^{yjk} \hspace{2em} (j, k = x, y, z) .
\eeq
The eigenvectors are
\beq
\bm{\alpha}_{\rm tr} (\mu = \pm 1) = \frac{1}{\sqrt{2}} \left(\begin{array}{r}
\mp 1 \\ 0 \\ i
\end{array}\right),
\hspace{2em}
\bm{\alpha}_{\rm tr} (\mu = 0) = 
\left(\begin{array}{r}
0 \\ 1 \\ 0
\end{array}\right) ,
\label{alpha_trans}
\eeq
where $\mu = -1, 0, 1$ denotes the $y$-spin eigenvalues. The phase of the $y$-polarized 
spinors and vectors, Eqs.~(\ref{chi_along_y}) and (\ref{alpha_trans}), is chosen such that they
correspond to the result of a finite rotation of the original $z$-polarized spinors 
and vectors, Eqs.(\ref{chi_along_z}) and (\ref{alpha_along_z}), by an angle
of $-\pi/2$ around the $x$-axis (i.e., the rotation that turns the positive $z$-axis 
into the positive $y$-axis, cf.~Fig.~\ref{fig:densities}). 
The vector-bispinor describing a spin-3/2 particle with transverse 
polarization is then given by the general formula Eq.~(\ref{vector_bispinor}), 
with the vectors and bispinors replaced by those with transverse polarization,
and the summation extending over the corresponding transverse polarization labels.

With the transversely polarized spinors we can now compute the $\pi N B$ vertex 
functions for transverse nucleon and baryon polarization. In the case of the intermediate 
nucleon $(B = N)$ \cite{Granados:2015lxa} the vertex for transverse polarization is given by an expression
analogous to the one for longitudinal polarization, Eq.~(\ref{vertex_nucleon_longit}), 
\be
\Gamma_{\pi N N, {\rm tr}} (y, \widetilde{\bm{k}}_T; \tau, \tau_1)
&=& \frac{2i g_A M_N}{F_\pi \sqrt{\bar{y}}}
\left[ y M_N \, S_{\rm tr}^z (\tau, \tau_1) + 
\widetilde{\bm{k}}_T \cdot \bm{S}_{\rm tr, T} (\tau, \tau_1) 
\right] ,
\label{vertex_restframe_y}
\ee
in which $S_{\rm tr}^z$ and $\bm{S}_{{\rm tr}, T} = (S_{\rm tr}^x, S_{\rm tr}^y)$ are
the components of the 3-vector characterizing the spin transition between transversely 
polarized nucleon states,
\be
S^i_{\rm tr} (\tau, \tau_1) &\equiv& \chi^\dagger_{\rm tr} (\tau)
({\textstyle\frac{1}{2}} \sigma^i ) \chi_{\rm tr} (\tau_1) 
\hspace{2em} (i = x, y, z).
\label{S_vector_y}
\ee
Note that now the $y$-component is diagonal,
\beq
S_{\rm tr}^y (\tau, \tau_1) 
\;\; = \;\; \tau \; \delta(\tau, \tau_1) ,
\eeq
while the $z$ and $x$ components have off-diagonal terms.
In the case of intermediate isobar ($B = \Delta$), the vertex for transverse 
polarization is given by an expression analogous to the one for longitudinal 
polarization, Eqs.~(\ref{vertex_delta_restframe_1}),
\beq
\Gamma_{\pi N \Delta, {\rm tr}} (y, \widetilde{\bm{k}}_T; \tau, \tau_1)
\;\; = \;\; \sum_{\mu\tau'} C^{3/2, \tau}_{1, \mu;\; 1/2, \tau'} \;
\bm{\alpha}_{\rm tr}(\mu) \cdot 
\bm{\mathcal{P}}_{\rm tr} (y, \widetilde{\bm{k}}_T, \tau', \tau_1) ,
\label{vertex_delta_transverse_1}
\eeq
in which $\bm{\alpha}_{\rm tr}$ is the transverse polarization 3-vector defined in 
Eq.~(\ref{alpha_trans}), and $\bm{\mathcal{P}}_{\rm tr} \equiv (\mathcal{P}^z_{\rm tr}, 
\bm{\mathcal{P}}_{{\rm tr}, T})$ is given by an
expression analogous to Eq.~(\ref{vertex_delta_restframe_2}),
\be
\left.
\begin{array}{cr}
\mathcal{P}^z_{\rm tr}
\\[3ex]
\bm{\mathcal{P}}_{{\rm tr}, T}
\end{array}
\right\} (y, \widetilde{\bm{k}}_T, \tau', \tau_1) \;
&=& \; \frac{\sqrt{2} g_{\pi N\Delta}}{M_N \bar y^{3/2}}
\left\{ \bar{M} \; \delta (\tau', \tau_1) + i \widetilde{\bm{k}}_T \cdot 
[ \bm{e}_z \times \bm{S}_{{\rm tr}, T}(\tau', \tau_1) ] \right\}
\; \left\{
\begin{array}{cr} 
\displaystyle
\frac{\widetilde{\bm{k}}_T^2}{2 M_\Delta} - M_d
\\[2ex]
- \widetilde{\bm{k}}_T 
\end{array}\right\} ,
\label{vertex_delta_transverse_2}
\ee
where now $\bm{S}_{{\rm tr}, T}$ is the transverse ($x, y$) component of the
3-vector Eq.~(\ref{S_vector_y}).

The $N \rightarrow \pi B$ light-front wave function of the nucleon for transverse nucleon 
and baryon polarization is then given by Eq.~(\ref{psi_restframe}), and its coordinate 
representation by Eq.~(\ref{psi_coordinate}), in the same way as for longitudinal polarization,
\be
\Psi_{N \rightarrow \pi B, \; {\rm tr}} (y, \widetilde{\bm{k}}_T; \tau, \tau_1)
&\equiv& 
\frac{\Gamma_{\pi N B, \; {\rm tr}} (y, \widetilde{\bm{k}}_T; \tau, \tau_1)}
{\Delta\mathcal{M}^2(y, \widetilde{\bm{k}}_T)} ,
\label{psi_restframe_y}
\\[1ex]
\Phi_{N \rightarrow \pi B, \; {\rm tr}} (y, \bm{r}_T, \sigma, \sigma_1) 
&\equiv&
\int \frac{d^2\widetilde{k}_{T}}{(2\pi)^2} \; 
e^{i \bm{\widetilde{k}}_T \cdot \bm{r}_T} \; 
\Psi_{N \rightarrow \pi B, \; {\rm tr}} (y, \bm{\widetilde{k}}_T; \sigma, \sigma_1) 
\hspace{2em} (B = N, \Delta).
\label{phi_tr}
\ee
The general decompositions Eq.~(\ref{coordinate_nucleon}) (for the nucleon)
and Eqs.~(\ref{R_z})-(\ref{R_T}) (for the isobar) apply to the transversely polarized
coordinate-space wave function as well, as they rely only on the functional dependence 
of the momentum-space wave function on $\widetilde{\bm{k}}_T$, not on the specific 
form of the spin structures. This allows us to express the transversely polarized
coordinate-space wave function in terms of the radial wave functions for the
longitudinally polarized system, using only the algebraic relations between
the spin structures. We quote only the expressions for the components with
initial transverse nucleon spin $\tau_1 = + {\textstyle\frac{1}{2}}$, as they
are sufficient to calculate the peripheral densities (cf.\ Sec.~\ref{sec:densities}).
For the intermediate nucleon ($B = N$) we obtain \cite{Granados:2015lxa}
\be
\Phi_{N \rightarrow \pi N, \; {\rm tr}} (y, \, \bm{r}_T, \, 
\tau = +{\textstyle\frac{1}{2}}, \, \tau_1 = +{\textstyle\frac{1}{2}}) 
&=&  \phantom{-}\sin\alpha \, U_1 ,
\label{phi_radial_1}
\\[1ex]
\Phi_{N \rightarrow \pi N, \; {\rm tr}} (y, \, \bm{r}_T, \, 
\tau = -{\textstyle\frac{1}{2}}, \, \tau_1 = +{\textstyle\frac{1}{2}}) &=& 
- U_0 \; + \; \cos\alpha \, U_1 
\label{phi_radial_2}
\\[1ex]
&& [U_{0,1} \; \equiv \; U_{0,1}(y, r_T)] ,
\nonumber
\ee
where $\alpha$ denotes the angle of the transverse vector $\bm{r}_T$ with respect to the
$x$-axis,
\be
\bm{r}_T &=& \; (r_T \cos\alpha, \, r_T \sin\alpha) .
\label{r_from_alpha}
\ee
For the intermediate isobar ($B = \Delta$) we obtain
\begin{eqnarray}
\Phi_{N \rightarrow \pi \Delta, \; {\rm tr}}
({\bm r}_T, y, \tau = +{\textstyle\frac{3}{2}},\tau_1 = +{\textstyle\frac{1}{2}})
&=& \phantom{-}
\frac{i}{\sqrt{2}}\left[V_0+W_0+(V_1+W_1)\cos{\alpha}+\frac{W_2}{2}\cos{2\alpha}\right] ,
\label{phi_radial_delta_1}
\\
\Phi_{N \rightarrow \pi \Delta, \; {\rm tr}}
({\bm r}_T, y, \tau=+{\textstyle\frac{1}{2}},\tau_1=+{\textstyle\frac{1}{2}})
&=&-\frac{i}{\sqrt{6}}\left[(V_1+2W_1)\sin{\alpha} + \frac{3W_2}{2}\sin{2\alpha}\right] ,
\label{phi_radial_delta_2}
\\
\Phi_{N \rightarrow \pi \Delta, \; {\rm tr}}
({\bm r}_T, y, \tau=-{\textstyle\frac{1}{2}},\tau_1=+{\textstyle\frac{1}{2}})
&=& \phantom{-}
\frac{i}{\sqrt{6}}\left[V_0+W_0+(V_1-W_1)\cos{\alpha} - \frac{3W_2}{2}\cos{2\alpha}\right] ,
\\
\Phi_{N \rightarrow \pi \Delta, \; {\rm tr}}
({\bm r}_T, y, \tau=-{\textstyle\frac{3}{2}},\tau_1=+{\textstyle\frac{1}{2}})
&=&-\frac{i}{\sqrt{2}}\left[V_1\sin{\alpha} - \frac{W_2}{2}\sin{2\alpha}\right]
\label{phi_radial_delta_4}
\\[1ex]
&& [V_{0,1} \; \equiv \; V_{0,1}(y, r_T), 
W_{0,1,2} \; \equiv \; W_{0,1,2}(y, r_T)] .
\nonumber
\end{eqnarray}

It is straightforward to compute the transverse densities in terms of the transversely 
polarized light-front wave functions. A simple result is obtained for the left and right current
densities, Eq.~(\ref{rho_left_right}):
\be
\left. 
\begin{array}{l}
\rho_{\rm left}^V (b) \\[2ex] \rho_{\rm right}^V (b) 
\end{array}
\right\}
&=& 
\frac{C_N}{4\pi} \int_0^1 \! \frac{dy}{y \bar y^3} 
\; \left| \Phi_{N \rightarrow \pi N, \; {\rm tr}} (y, \; \mp r_T \bm{e}_x; \;
\tau = -{\textstyle\frac{1}{2}}, \; \tau_1 = {\textstyle\frac{1}{2}}) \right|^2 ,
\label{rho_left_right_transverse}
\\[3ex]
\left. 
\begin{array}{l}
\rho_{\rm left}^V (b) \\[2ex] \rho_{\rm right}^V (b) 
\end{array}
\right\}
&=& 
\frac{C_\Delta}{6\pi} \int_0^1 \! \frac{dy}{y \bar y^3}
\; \sum_{\tau = -1/2, \; 3/2}
\; \left| \Phi_{N \rightarrow \pi \Delta, \; {\rm tr}} 
(y, \; \mp r_T \bm{e}_x; \; \tau, \; \tau_1 = {\textstyle\frac{1}{2}}) \right|^2
\label{rho_left_right_transverse_delta}
\\[2ex]
&& [r_T = b/(1 - y)] .
\nonumber
\ee
If we substitute in Eqs.~(\ref{rho_left_right_transverse}) and
(\ref{rho_left_right_transverse_delta})
the explicit expressions for $\Phi_{N \rightarrow \pi B, \; {\rm tr}}$
in terms of the radial wave functions, 
Eqs.~(\ref{phi_radial_1})-(\ref{phi_radial_2})
and Eqs.~(\ref{phi_radial_delta_1})-(\ref{phi_radial_delta_1}),
and note that the ``left'' and ``right'' points $\bm{r}_T = \mp r_T \bm{e}_x$
correspond to $\alpha = \pi$ and $0$, respectively [cf.~Eq.~(\ref{r_from_alpha})],
we recover the expressions obtained previously with longitudinally polarized nucleon
states, Eqs.~(\ref{rho_left_right_longitudinal_delta}) and
(\ref{rho_left_right_longitudinal}). It shows that the two calculations, using transversely
or longitudinally polarized nucleon states, give the same results for the densities
if the $N \rightarrow \pi B$ light-front wave functions are related as described above. 
We note that this correspondence is made possible by the use of light-front helicity 
states, which allow one to relate longitudinally and transversely polarized states 
in a simple manner by a rest-frame spin rotation.

The transversely polarized expressions Eqs.~(\ref{rho_left_right_transverse}) and
(\ref{rho_left_right_transverse_delta}) have several remarkable properties.
First, the densities are diagonal in the external nucleon transverse spin ($\tau_2 = \tau_1$).
They describe a sum of contributions of individual $\pi B$ states (``orbitals'') in
the nucleon's wave function, which have definite sign and permit a simple probabilistic
interpretation. Second, one sees that only specific intermediate spin states contribute
to the left and right current densities on the $x$-axis. Parity conservation
dictates that the $\pi B$
system be in a state with odd orbital angular momentum $L$, such that $(-1)^L = -1$.
Angular momentum conservation requires that the spins of the initial nucleon and the
intermediate baryon and the orbital angular momentum be related by the vector coupling
rule, which leaves only the state with $L = 1$ for both $B = N, \Delta$. 
The sums in Eqs.~(\ref{rho_left_right_transverse}) and 
(\ref{rho_left_right_transverse_delta}) run over the intermediate baryon's 
$y$-spin $\tau$, and thus effectively over the $y$-projection of the orbital
angular momentum,
\beq
L^y \;\; = \;\; \tau_1 - \tau \;\; = \;\; {\textstyle\frac{1}{2}} - \tau , 
\eeq
which can take on the values $L^y = (-1, 0, 1)$. Further simplification comes about 
because the wave function of the $\pi B$ state with $L^y = 0$ vanishes on the
$x$-axis (i.e., in the direction perpendicular to the spin quantization axis),
as seen in the explicit expressions Eq.~(\ref{phi_radial_1}) (for $B = N$) and 
Eq.~(\ref{phi_radial_delta_2}) (for $B = \Delta$), which are zero for $\alpha = 0$
or $\pi$. Altogether this leaves only the following intermediate states to
contribute to the densities:
\beq
\begin{array}{|r|r|r|r|}
\hline
\multicolumn{4}{|c|}{N(\tau_1) \rightarrow \pi B(\tau)} \\[.5ex]
\hline
\hline
 B     & \tau_1 & \tau & L^y \\[.5ex]
\hline
\hline
N & {\textstyle\frac{1}{2}} & -{\textstyle\frac{1}{2}} & 1 \\[.5ex]
\hline
\Delta & {\textstyle\frac{1}{2}} & -{\textstyle\frac{1}{2}} & 1 \\[.5ex]
           & {\textstyle\frac{1}{2}} &  {\textstyle\frac{3}{2}} & -1 \\[.5ex]
\hline
\end{array}
\label{quantum_numbers}
\eeq
Thus a single orbital accounts for the transverse densities generated by the nucleon intermediate
state, and two orbitals for the ones generated by the isobar. This simple structure makes
the transversely polarized representation a convenient framework for studying the 
properties of the peripheral transverse densities in chiral EFT.
\subsection{Mechanical interpretation}
%
%
\begin{figure}
\includegraphics[width=.6\textwidth]{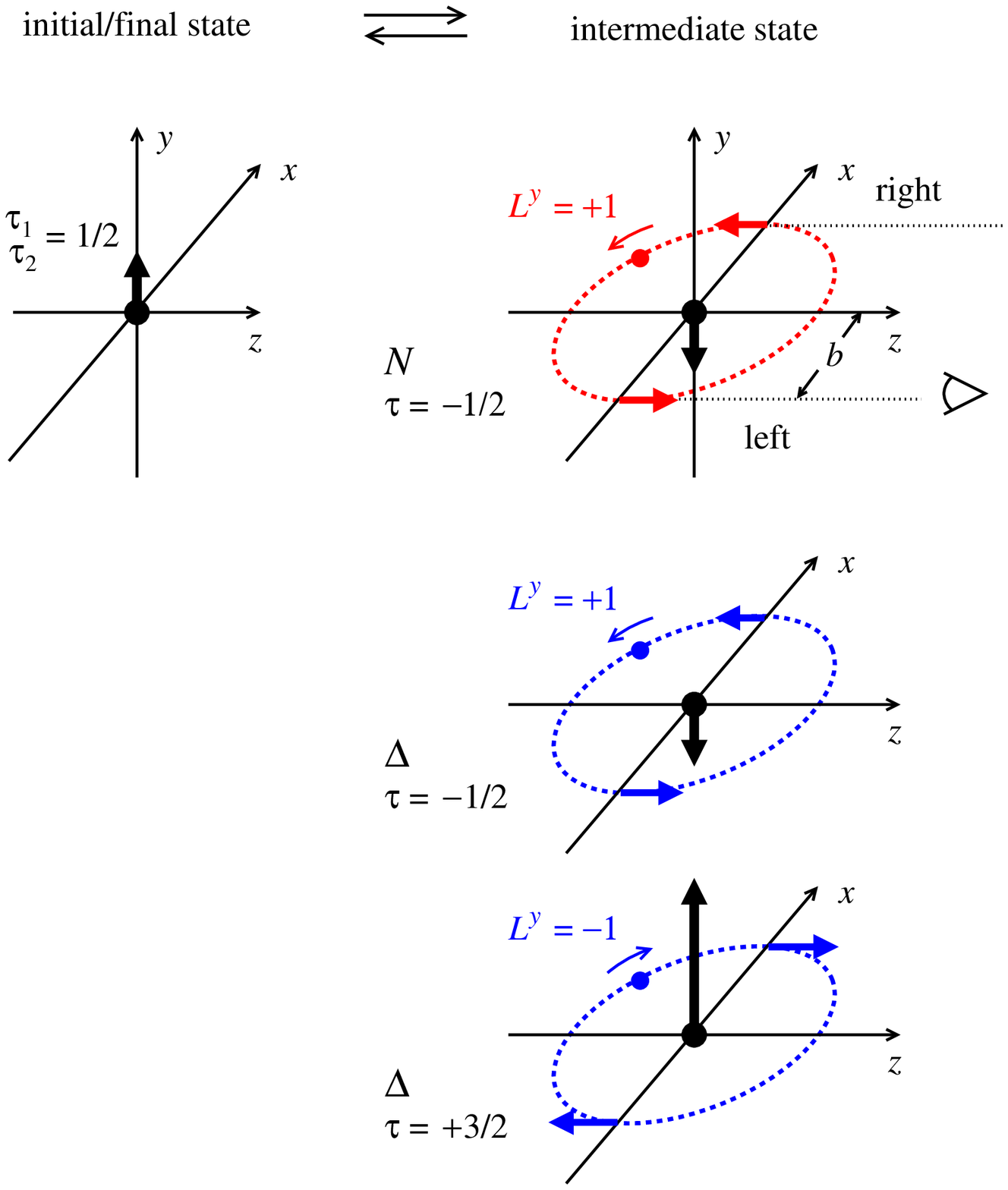}
\caption[]{Quantum-mechanical picture of the peripheral transverse densities 
in chiral EFT with $\Delta$ isobars. The bare nucleon with $y$-spin projection
$\tau_1 = +1/2$ in the rest frame (left) makes a transition to a pion-baryon intermediate 
state with baryon $y$-spin projection $\tau$ and pion angular momentum projection $L^y$ (right), 
and back to a bare state with $\tau_2 = +1/2$.
Top: $\pi N$ intermediate state with $\tau = -1/2$ and $L^y = +1$. Middle, bottom: $\pi \Delta$
intermediate states with $\tau = -1/2$ and $L^y = +1$, and $\tau = +3/2$ and $L^y = -1$. The peripheral 
left/right transverse densities $\rho^V_{\rm left/right}(b)$ describe the $J^+$ current at the
positions $\bm{b} = \mp b \bm{e}_x$ (on the left and right $x$-axis, viewed from $z = +\infty$),
which arises from the convection current produced by the charged pion. The spin-independent and
-dependent densities, $\rho_1^V (b)$ and $\widetilde\rho_2^V (b)$, describe the left-right average
and left-right asymmetry of the current (cf.\ Fig.~\ref{fig:densities}). 
\label{fig:mech}}
\end{figure}
Our results for the peripheral transverse densities with intermediate $\Delta$ isobars can be explained 
in the quantum-mechanical picture described in Refs.~\cite{Granados:2015lxa,Granados:2015rra}.
One adopts a first-quantized view of chiral EFT and follows the evolution of the chiral processes
in light-front time. In this view the initial ``bare'' nucleon in the rest frame, with $y$-spin projection
$\tau_1 = +1/2$, makes a transition to a pion-baryon (nucleon or isobar) intermediate state
with baryon $y$-spin projection $\tau$ and pion angular momentum projection $L^y$ (right), 
and back to the bare final nucleon with $\tau_2 = +1/2$ (see Fig.~\ref{fig:mech}). 
The allowed quantum numbers of the intermediate states are the ones listed in Eq.~(\ref{quantum_numbers}). 
The peripheral transverse densities then arise as the result of the convection current carried by 
the charged pion in the intermediate state. In particular, the left and right densities, 
$\rho^V_{\rm left}(b)$ and $\rho^V_{\rm right}(b)$, are given by
the $J^+$ current at the positions $\bm{b} = \mp b \bm{e}_x$ (on the left and right $x$-axis, 
viewed from $z = +\infty$, cf.~Fig.~\ref{fig:densities}a). 

In a plane-wave state the current carried by a charged pion is proportional to its 4-momentum, 
$\langle \pi (k)|J^+| \pi(k) \rangle = 2 Q_\pi k^+ = 2 Q_\pi [E_\pi(\bm{k}) + k^z]$, where 
$Q_\pi = \pm 1$ is the pion charge. The typical pion momenta
in chiral EFT processes are $|\bm{k}| = O(M_\pi)$, and the energies are $E_\pi = O(M_\pi)$, which 
means that the motion of the pion is essentially relativistic. For such pions the plus current is 
generally much larger if $k^z > 0$ than if $k^z < 0$, i.e., if the pion moves in the positive rather 
than the negative $z$-direction. Now the peripheral pion in the intermediate states of the 
EFT processes considered here is outside of the range of the EFT interactions
(which are pointlike on the scale $M_\pi$), so that its current is effectively that of a free pion
(see Figure~\ref{fig:mech}). One therefore concludes that the current on the side where the pion moves 
in the positive $z$-direction is much larger than on the side where it moves in the opposite direction. 
This basic fact explains the pattern of left and right transverse densities observed in our calculation 
with transversely polarized states.

In the $\pi N$ intermediate state the only allowed transverse spin quantum numbers 
are $\tau = -1/2$ and $L^y = +1$ (see Fig.~\ref{fig:mech}). In this state the pion plus current is larger on 
the left than on the right, so that
\beq
|\rho_{{\rm left}, N}(b)| \;\; \gg \;\; |\rho_{{\rm right}, N}(b)| .
\eeq
This is indeed observed in the numerical results of Fig.~\ref{fig:densb}b. The large left-to-right ratio 
of $\sim 4-10$ shows that the motion of the pion is highly relativistic, because for a non-relativistic 
pion the left-to-right ratio would be close to unity,
\beq
\frac{|\rho_{{\rm left}, N}(b)|}{|\rho_{{\rm right}, N}(b)|} \;\; = \;\; 1 + O(v),
\eeq
where $v \ll 1$ is the characteristic velocity of the pion. In the $\pi \Delta$ intermediate state 
the allowed transverse spin quantum numbers are 
$\tau = -1/2$ and $L^y = +1$, and $\tau = +3/2$ and $L^y = -1$. From the $\pi N\Delta$
vertex for transverse polarization, Eq.~(\ref{vertex_delta_transverse_1}), one can see that the 
spin wave function of the intermediate state is
\beq
\frac{1}{\sqrt{6}} |\tau = -{\textstyle\frac{1}{2}}, L^y = +1\rangle
\; + \; 
\frac{1}{\sqrt{2}} |\tau = {\textstyle\frac{3}{2}}, L^y = -1\rangle \; + \; \ldots ,
\eeq
where we omitted the components that do not contribute to the current at $\bm{b} = \pm b \bm{e}_x$.
The configuration with $L^y = -1$ is 3 times more likely than the one with $L^y = 1$ and dominates 
the transverse densities. In the $\pi\Delta$ intermediate states the pion therefore moves predominantly 
with $L^y = -1$, and the pion plus current is larger on the right than on the left
\beq
|\rho_{\rm left, \Delta}(b)| \;\; \ll \;\; |\rho_{\rm right, \Delta}(b)| .
\eeq
This is again observed in the numerical results of Fig.~\ref{fig:densb}b.

The sign of the transverse densities can be deduced by considering the isospin structure of the $\pi B$
intermediate states. Let us consider the case that the external nucleon is a proton ($p$). In $\pi N$ 
intermediate states, the only configuration with a charged pion is $\pi^+ n$, and therefore
\beq
\rho_{\rm left, N}(b), \;  \rho_{\rm right, N}(b) \;\; > \;\; 0. 
\eeq
In $\pi\Delta$ intermediate states the isospin structure is [cf.~Eq.~(\ref{lag_delta})]
\beq
\frac{1}{\sqrt{2}}|\Delta^{++} \pi^- \rangle \; + \; 
\frac{1}{\sqrt{6}} |\Delta^0 \pi^+ \rangle \; + \; 
\frac{1}{\sqrt{3}} |\Delta^+ \pi^0 \rangle ,
\eeq
so that the configurations with $\pi^-$ are 3 times more likely than the ones with $\pi^+$.
The peripheral current is therefore dominated by the negatively charged pion, and one has
\beq
\rho_{\rm left, \Delta}(b), \;  \rho_{\rm right, \Delta}(b) \;\; < \;\; 0 .
\eeq
This pattern is seen in the numerical results of Fig.~\ref{fig:densb}b.

In sum, the peripheral transverse densities obtained from chiral EFT can naturally be explained
in terms of the orbital motion of the peripheral pion and the isospin structure of the intermediate 
state in the quantum-mechanical picture. The large left-right asymmetry attests to the essentially 
relativistic motion of pions, which is a fundamental property of 
chiral dynamics \cite{Granados:2015lxa,Granados:2015rra}. The $\Delta$ isobar introduces intermediate 
states in which the orbital motion of the peripheral pion is ``reversed'' compared to the intermediate
state with the nucleon, and in which the pion has ``reversed'' charge, resulting in a rich structure.
The double reversal explains why the $N$ and $\Delta$ contributions compensate each other in the
spin-independent density $\rho_1^V$ but add in the spin-dependent density $\widetilde\rho_2^V$.
We emphasize that the mechanical picture presented here is derived from an \textit{exact rewriting} 
of LO relativistic chiral EFT and has an objective dynamical content, which distinguishes it from 
phenomenological pion cloud models of nucleon structure.
\section{Large-$N_c$ limit of transverse densities}
\label{sec:largenc}
\subsection{Light-front wave functions at large $N_c$}
\label{subsec:largenc}
We now want to study how the light-front representation of chiral EFT with $\Delta$ isobars 
behaves in the large-$N_c$ limit of QCD. This exercise will further elucidate the structure 
of the wave function overlap formulas and provide a crucial test for the isobar results.
It also allows us to re-derive the $N_c$-scaling relations for the peripheral densities in a
simple manner and interpret them in the context of the mechanical picture of Sec.~\ref{sec:mechanical}.

The limit of a large number of colors in QCD represents a general method for relating properties of mesons 
and baryons to the microscopic theory of strong interactions \cite{'tHooft:1973jz,Witten:1979kh}. 
While even at large $N_c$ the dynamics remains complex and cannot be solved exactly, the scaling behavior of 
meson and baryon properties with $N_c$ can be established on general grounds and provides interesting insights 
and useful constraints for phenomenology. It is found that the masses of 
low-lying mesons (including the pion) scale as $O(N_c^0)$, while those of baryons scale as $O(N_c)$ 
for states with spin/isospin $O(N_c^0)$. The basic hadronic size of mesons and baryons scales as
$O(N_c^0)$ and remains stable in the large-$N_c$ limit. Baryons at large $N_c$ thus are heavy objects
of fixed spatial size, whose external motion (in momentum and spin/isospin) can be described classically 
and is governed by inertial parameters of order $O(N_c)$ (mass, moment of inertia). The $N$ and $\Delta$ 
appear as rotational states of the classical body with spin/isospin $S = T = 1/2$ and $S = T = 3/2$,
respectively, and the mass splitting is $M_\Delta - M_N = O(N_c^{-1})$. Further scaling relations are
obtained for the transition matrix elements of current operators between meson and baryon states,
as well as the meson-meson and meson-baryon couplings; see Ref.~\cite{Jenkins:1998wy} for a review. 
The relations are model-independent and can be derived in many different ways, e.g.\ 
using diagrammatic techniques \cite{Witten:1979kh},
group-theoretical methods \cite{Dashen:1993jt,Dashen:1994qi}, large-$N_c$ quark models
\cite{Karl:1984cz,Jackson:1985bn}, or the soliton picture of baryons \cite{Adkins:1983ya,Zahed:1986qz}.

The large-$N_c$ limit can be combined with an EFT descriptions of strong interactions in terms of 
meson-baryon degrees of freedom, valid in special parametric regions. The application of $N_c$-scaling
relations to chiral EFT has been studied using formal methods, focusing on the interplay of the 
limits $M_\pi \rightarrow 0$ and $N_c \rightarrow \infty$ in the calculation of spatially integrated 
quantities (charges, RMS radii) \cite{CalleCordon:2012xz,Cohen:1992uy,Cohen:1996zz}. 
In the present study of spatial densities, following the approach of Ref.~\cite{Granados:2013moa}, 
we impose the $N_c$-scaling relations for the $N$ and $\Delta$ masses, and their 
couplings to pions, and calculate the resulting scaling behavior of the peripheral transverse 
densities at distances $O(M_\pi^{-1})$. This is consistent with our general philosophy of
regarding the pion mass as finite but small on the hadronic scale, $M_\pi \ll \mu_{\rm had}$, and 
applying chiral EFT to peripheral hadron structure at distances $M_\pi^{-1} \gg \mu_{\rm had}^{-1}$. 

The general $N_c$-scaling of the nucleon's isovector transverse charge and magnetization densities 
at non-exceptional distances was established in Ref.~\cite{Granados:2013moa} and is of the form
\beq
\rho_1^V (b) \;\; = \;\; O(N_c^0),
\hspace{2em}
\widetilde{\rho}_2^V (b) \;\; = \;\; O(N_c)
\hspace{2em}
[b = O(N_c^0)].
\label{rho_largen_general}
\eeq
These relations follow from the known $N_c$-scaling of the total charge and magnetic moment, and the 
scaling of the nucleon size as $O(N_c^0)$. In chiral EFT we consider the densities at peripheral 
distances $b \sim M_\pi^{-1}$. Because $M_\pi = O(N_c^0)$, the chiral region remains stable 
in the large-$N_c$ limit, 
\beq
b \; \sim \; M_\pi^{-1} \; = \; O(N_c^0).
\label{b_chiral_largen}
\eeq
The densities generated by chiral dynamics thus represent a distinct component of the nucleon's spatial
structure even in the large-$N_c$ limit. Using the dispersive representation of the transverse densities
it was shown in Ref.~\cite{Granados:2013moa} that the LO chiral EFT results obey the general $N_c$-scaling 
laws Eq.~(\ref{rho_largen_general}) if the isobar contribution is included. 

We now want to investigate how the $N_c$-scaling relations for the transverse densities arise 
in the light-front representation of chiral EFT. To this end we consider the $N \rightarrow \pi B$ 
light-front wave functions of Sec.~\ref{sec:lightfront} with masses scaling as
\beq
M_N, M_\Delta \; = \; O(N_c),
\hspace{3em}
M_\Delta - M_N \; = \; O(N_c^{-1}) ,
\hspace{3em}
M_\pi \; = \; O(N_c^0) .
\eeq
The $N_c$-scaling of the $\pi NN$ coupling follows from $g_A = O(N_c)$ and
$F_\pi = O(N_c^{1/2})$ and is
[cf.\ Eq.~(\ref{axial_pseudoscalar})]
\beq
g_{\pi NN} \;\; \equiv \;\; \frac{g_A M_N}{F_\pi} \;\; = \;\; O(N_c^{3/2}).
\label{g_pinn_nc}
\eeq
The $\pi N\Delta$ coupling in the large-$N_c$ limit is related to the 
$\pi NN$ coupling by \cite{Adkins:1983ya}
\beq
g_{\pi N\Delta} \;\; = \;\; \frac{3}{2} \; g_{\pi NN} \hspace{2em} (N_c \rightarrow \infty )
\label{g_pindelta_nc}
\eeq
and follows the same $N_c$-scaling. 
When calculating the densities at chiral distances, Eq.~(\ref{b_chiral_largen}), 
the typical pion momenta in the nucleon rest frame are $|\bm{k}| \sim M_\pi = O(N_c^0)$, and the
light-front wave functions are evaluated in the region
\beq
\left.
\begin{array}{rclcl}
y &\sim& M_\pi / M_N &=& O(N_c^{-1})
\\[1ex] 
|\widetilde{\bm{k}}_T| &\sim& M_\pi &=& O(N_c^0) 
\\[1ex] 
|\bm{r}_T| &\sim& M_\pi^{-1} &=& O(N_c^0) 
\end{array}
\right\} .
\label{nc_region}
\eeq
For such configurations the invariant mass denominator Eq.~(\ref{invariant_mass_restframe}) 
is of the order
\beq
\Delta\mathcal{M}^2_{N \rightarrow \pi B} \;\; = \;\; O(N_c)
\hspace{2em} (B = N, \Delta).
\eeq
The $N_c$-scaling of the vertex function for general nucleon and baryon helicities follows
from Eqs.~(\ref{g_pinn_nc}) and (\ref{g_pindelta_nc}),
\beq
\Gamma_{\pi N B} \;\; = \;\; O(N_c^{3/2}) \hspace{2em} (B = N, \Delta).
\eeq
Altogether the $N_c$-scaling of the light-front wave functions in the region
Eq.~(\ref{nc_region}) is
\beq
\left.
\begin{array}{lcl}
\Psi_{N \rightarrow \pi B} (y, \widetilde{\bm{k}}_T, \ldots) &=& O(N_c^{1/2}) 
\\[1ex]
\Phi_{N \rightarrow \pi B} (y, \bm{r}_T, \ldots) &=& O(N_c^{1/2}) 
\end{array}
\right\}
\hspace{2em} (B = N, \Delta).
\label{nc_wf_general}
\eeq
Equation~(\ref{nc_wf_general}) represents the ``maximal'' scaling behavior for general 
nucleon and baryon helicities. Combinations of helicity components can have a lower
scaling exponent due to cancellations of the leading term.

We now want to inspect the $N_c$-scaling of the individual helicity structures and compare 
the nucleon and isobar wave functions. It is straightforward to establish the scaling behavior
of the radial wave functions from the explicit expressions in 
Eqs.~(\ref{radial_nucleon}) and (\ref{radial_delta}). In the case of the intermediate
nucleon ($B = N$), the two radial functions are of the same order in $N_c$,
\beq
U_0, U_1 \;\; = \;\; O(N_c^{1/2}) \hspace{2em} [y = O(N_c^{-1}), \; r_T = O(N_c^0)] .
\label{U_largen}
\eeq
In the case of the isobar ($B = \Delta$), the radial functions $V_0$ and $W_1$ are
leading in $N_c$, while $V_1, W_0$ and $W_2$ are subleading,
\beq
\left.
\begin{array}{lcl}
V_0, W_1 &=& O(N_c^{1/2})  
\\[1ex]
V_1, W_0, W_2 &=& O(N_c^{-1/2})
\end{array}
\right\}
\hspace{2em} [y = O(N_c^{-1}), \; r_T = O(N_c^0)] .
\eeq
Furthermore, the transverse mass Eq.~(\ref{M_T_def}), which governs the 
radial dependence of the wave functions, becomes the same for $B = N$ and $\Delta$
in the large-$N_c$ limit,
\beq
M_{T, \Delta}^2(y) \;\; \approx \;\; M_{T, N}^2(y) \;\; \approx \;\;
M_\pi^2 + y^2 M_N^2  \; + \;\; O(N_c^{-1})
\hspace{2em} (N_c \rightarrow \infty) .
\eeq
This happens because the term proportional to the $N$-$\Delta$ mass splitting 
in Eq.~(\ref{M_T_def}) scales as 
$y (M_\Delta^2 - M_N^2) = y (M_\Delta + M_N)(M_\Delta - M_N) = O(N_c^{-1})$ 
and is suppressed at large $N_c$. As a result the leading light-front wave functions
for the intermediate $N$ and $\Delta$ at large $N_c$ have pairwise identical radial dependence.
Using also the relations between the couplings, Eq.~(\ref{g_pindelta_nc}), we obtain
\beq
\left.
\begin{array}{l}
V_0 (y, r_T)
\\[1ex]
W_1 (y, r_T)
\end{array}
\right\} 
\; \approx \; 
\frac{3}{\sqrt{2}}
\left\{
\begin{array}{l}
U_0 (y, r_T)
\\[1ex]
U_1 (y, r_T)
\end{array}
\right\}
\hspace{2em}
[y = O(N_c^{-1}), \; r_T = O(N_c^0)] .
\label{n_delta_radial_largen}
\eeq
The relations Eqs.~(\ref{U_largen})-(\ref{n_delta_radial_largen}) together embody the $N_c$-scaling
of the full set of chiral light-front wave functions. They naturally explain the patterns observed 
in the intermediate $N$ and $\Delta$ wave functions at finite $N_c$ (i.e., calculated with the 
physical baryon masses and couplings) shown in Figure~\ref{fig:wfy}a and b.

The $N_c$-scaling relations for the light-front wave functions now allow us to explain the scaling 
behavior of the transverse densities in a simple manner. Using 
Eqs.~(\ref{U_largen})-(\ref{n_delta_radial_largen}) in the overlap formulas,
Eqs.~(\ref{rho_overlap_n}) and (\ref{rho_overlap_delta}), we obtain for the spin-independent
density
\be
\rho_{1, N}^V (b) \; = \; O(N_c), 
\hspace{2em} \rho_{1, \Delta}^V (b) \; = \; O(N_c) .
\ee
The scaling exponents of the intermediate $N$ and $\Delta$ contributions {\it alone} is 
{\it larger} than that of the general scaling relation Eq.~(\ref{rho_largen_general}).
The correct $N_c$-scaling of the chiral EFT result is obtained by {\it combining} the
intermediate $N$ and $\Delta$ contributions. Using Eq.~(\ref{n_delta_radial_largen})
and the specific values of the isospin factors $C_N$ and $C_\Delta$, Eq.~(\ref{isospin_factors}),
the $O(N_c)$ terms cancel, and we obtain
\be
\rho_{1, N}^V (b) \; + \; 
\rho_{1, \Delta}^V (b) \; = \; O(N_c^{-1}), 
\label{rho_1_n_delta_nc}
\ee
in accordance with Eq.~(\ref{rho_largen_general}). For the spin-dependent density
we obtain
\be
\widetilde\rho_{2, N}^V (b) \; = \; O(N_c), 
\hspace{2em} \widetilde\rho_{2, \Delta}^V (b) \; = \; O(N_c), 
\ee
Here the $N$ and $\Delta$ contributions alone already show the correct scaling behavior
as required by Eq.~(\ref{rho_largen_general}). Using Eqs.~(\ref{n_delta_radial_largen})
and (\ref{isospin_factors}) one sees that in the large-$N_c$ limit the intermediate
$\Delta$ contribution is exactly $1/2$ times the intermediate $N$ one, so that the
combined result is $3/2$ times the $N$ contribution alone,
\be
\widetilde{\rho}_{2, \Delta}^V (b) &=& \frac{1}{2} \;
\widetilde{\rho}_{2, N}^V (b) \; + \; O(N_c^{-1}), 
\\[1ex]
\widetilde{\rho}_{2, N}^V (b) + \widetilde{\rho}_{2, \Delta}^V (b) 
&=& \frac{3}{2} \; \widetilde{\rho}_{2, N}^V (b) \; + \;
O(N_c^{-1}), 
\label{rho_2_tilde_n_delta_nc}
\ee
as found in the dispersive calculation of Ref.~\cite{Granados:2013moa}. Altogether, we see that
the chiral EFT results with the isobar reproduce the general $N_c$-scaling of the
transverse densities. In the mechanical picture of Sec.~\ref{sec:mechanical}, Eq.~(\ref{rho_1_n_delta_nc}) 
is realized by the cancellation of the currents produced by the peripheral pion 
in states with intermediate $N$ and $\Delta$; Eq.(\ref{rho_2_tilde_n_delta_nc}), by 
the addition of the same currents.
\subsection{Contact terms and off-shell behavior}
\label{subsec:contact}
We now want to evaluate the contact term contributions to the peripheral transverse densities
and verify that they, too, obey the general $N_c$-scaling laws. This exercise also exposes 
the effect of the off-shell ambiguity of the chiral EFT with isobars on the peripheral densities,
and shows to what extent the ambiguity can be constrained by $N_c$-scaling arguments.

Following the decomposition of the LO current matrix element in Sec.~\ref{subsec:current}, the
peripheral densities are given by the sum of the intermediate-baryon and contact 
terms,\footnote{In this section we use explicit labels ``interm'' and ``cont'' to denote
the intermediate-baryon and contact contribution to the densities; cf.~Footnote~\ref{foot:interm}.}
\beq
\left.
\begin{array}{lcl}
\rho_{1, B}^V(b) &=&  
\rho_{1, B}^V(b)_{\rm interm} \; + \; \rho_{1, B}^V (b)_{{\rm cont}}
\\[1ex]
\widetilde\rho_{2, B}^V(b) &=&  
\widetilde\rho_{2, B}^V(b)_{\rm interm} \; + \; \widetilde\rho_{2, B}^V (b)_{{\rm cont}}
\end{array}
\right\}
\hspace{2em} (B = N, \Delta) .
\eeq
The intermediate-baryon terms were expressed as light-front wave function overlap
in Sec.~\ref{subsec:overlap}, and their $N_c$-scaling studied in Sec.~\ref{subsec:largenc}.
The contact terms 
are computed by evaluating the Feynman integrals, Eqs.~(\ref{current_nucleon_contact}) and
(\ref{numerator_nucleon_contact}), and Eqs.~(\ref{current_delta_contact}) and
(\ref{numerator_delta_contact}), directly as four-dimensional integrals,
using the fact that they depend on the 4-momentum transfer $\Delta$
as the only external 4-vector \cite{Granados:2013moa}. We obtain
\be
\rho_{1, N}^V(b)_{\rm cont} &=& \frac{(1 - g_A^2) M_\pi^4}{12 \pi^3 F_\pi^2} \; R_{\rm cont}(M_\pi b) ,
\label{rho_1_contact_nucleon}
\\[1ex]
\rho_{1, \Delta}^V(b)_{\rm cont} &=& \frac{g_{\pi N\Delta}^2 M_\pi^4 (M_\Delta + M_N)^2}{108 \pi^3 
M_\Delta^2 M_N^2} \; R_{\rm cont}(M_\pi b) ,
\label{rho_1_contact_delta}
\\[2ex]
\widetilde\rho_{2, N}^V(b)_{\rm cont} &=& 0,
\label{rho_2_contact_nucleon}
\\[2ex]
\widetilde\rho_{2, \Delta}^V(b)_{\rm cont} &=& \frac{g_{\pi N\Delta}^2 M_\pi^5}{108 \pi^3 
M_\Delta^2 M_N} \; R_{\rm cont}^\prime (M_\pi b) ,
\label{rho_2_contact_delta}
\ee
where $R_{\rm cont}$ denotes the normalized loop integral introduced in Appendix B of 
Ref.~\cite{Granados:2013moa} ($\beta = M_\pi b$)
\be
R_{\rm cont}(\beta) &\equiv& \frac{1}{16} \left\{ 3 [K_0(\beta)]^2 - 4 [K_1(\beta)]^2 + [K_2(\beta)]^2 
\right\} ,
\\[1ex]
R_{\rm cont}^\prime (\beta) \; \equiv \; \frac{\partial}{\partial \beta} R_{\rm cont}(\beta) 
&=& \frac{1}{16} \left[ - 2 K_0(\beta) K_1(\beta) + 3 K_1(\beta) K_2(\beta) - K_2(\beta) K_3(\beta) 
\right] .
\ee
In deriving Eq.~(\ref{rho_1_contact_delta}) we have neglected terms $O(M_\pi^2/M_N^2)$ in the coefficient;
this approximation is consistent with both the chiral and the $1/N_c$ expansions, and the neglected terms
are numerically small.

We can now study the $N_c$-scaling of the contact terms. In the spin-independent density $\rho_1^V$ 
we observe that 
\be
\rho_{1, N}^V(b)_{\rm cont} &=& O(N_c) ,
\\[1ex]
\rho_{1, \Delta}^V(b)_{\rm cont}  &=& O(N_c) ,
\\[1ex]
\rho_{1, \Delta}^V(b)_{\rm cont} + \rho_{1, N}^V(b)_{\rm cont} \; &=& \; O(N_c^0).
\label{rho_1_contact_largen}
\ee
The $N$ and $\Delta$ contact terms are individually $O(N_c)$, as the intermediate-baryon terms, 
but they cancel each other in leading order, as it should be. This remarkable result comes about 
thanks to two circumstances: (a) in the $N$ contact term the piece proportional to $g_A^2 = O(N_c^2)$
dominates; this piece arises from the nucleon triangle diagram and has the same origin as the $\Delta$
contact term; (b) the coupling constants in the large-$N_c$ limit are related by
Eqs.~(\ref{g_pinn_nc}) and (\ref{g_pindelta_nc}). Note that the contact and intermediate-baryon terms 
are of the same order in $N_c$ in the spin-independent density,
\beq
\frac{\rho_{1, B}^V(b)_{\rm cont}}{\rho_{1, B}^V(b)_{\rm interm}} \;\; = \;\; O(N_c^0) 
\hspace{2em} (B = N, \Delta) .
\eeq
The cancellation between the nucleon and isobar contact terms, Eq.~(\ref{rho_1_contact_largen}), is 
thus essential in bringing about the correct $N_c$-scaling of the overall spin-independent densities.
In the spin-dependent density $\widetilde\rho_2^V$ there is no contribution from the nucleon contact term 
in LO of the chiral expansion (a non-zero contribution appears at NLO and scales as $O(N_c^{-1})$ and is 
therefore suppressed in large-$N_c$ limit \cite{Kubis:2000zd,Granados:2013moa}). The isobar contact 
term Eq.~(\ref{rho_2_contact_delta}) scales as 
\be
\widetilde\rho_{2, \Delta}^V(b)_{\rm cont}  &=& O(N_c^{-1}) ,
\ee
and is therefore of the same order in $N_c$ as the nucleon contact term, as is natural.
Both nucleon and isobar contact terms are therefore strongly suppressed relative to the 
intermediate-baryon terms in the spin-dependent density,
\beq
\frac{\widetilde\rho_{2, B}^V(b)_{\rm cont}}{\widetilde\rho_{2, B}^V(b)_{\rm interm}} \;\; = \;\; O(N_c^{-2}) 
\hspace{2em} (B = N, \Delta) .
\eeq
In sum, the full peripheral densities, including the contact terms, obey the same large-$N_c$ relations
as the intermediate-baryon contributions represented by the light-front wave function overlap integrals. 
In the spin-independent case this comes about due to a non-trivial cancellation between the nucleon and 
isobar contact terms.

%
%
\begin{figure}
\begin{tabular}{l}
\includegraphics[width=.48\textwidth]{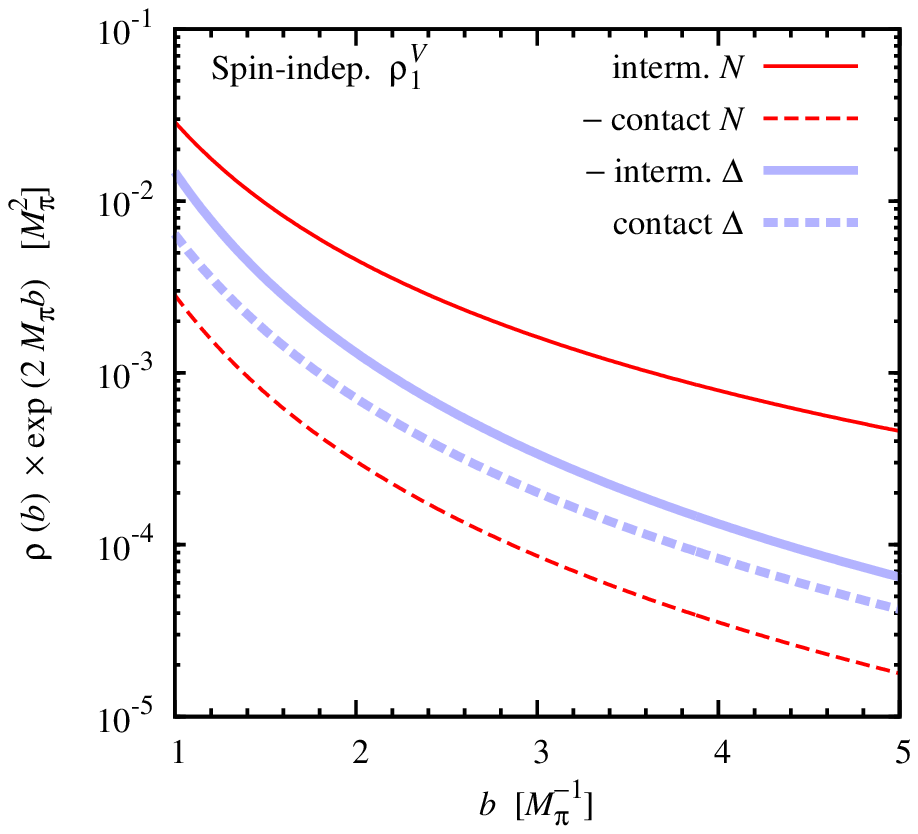}
\\[-4ex]
{\bf (a)}
\\[2ex]
\includegraphics[width=.48\textwidth]{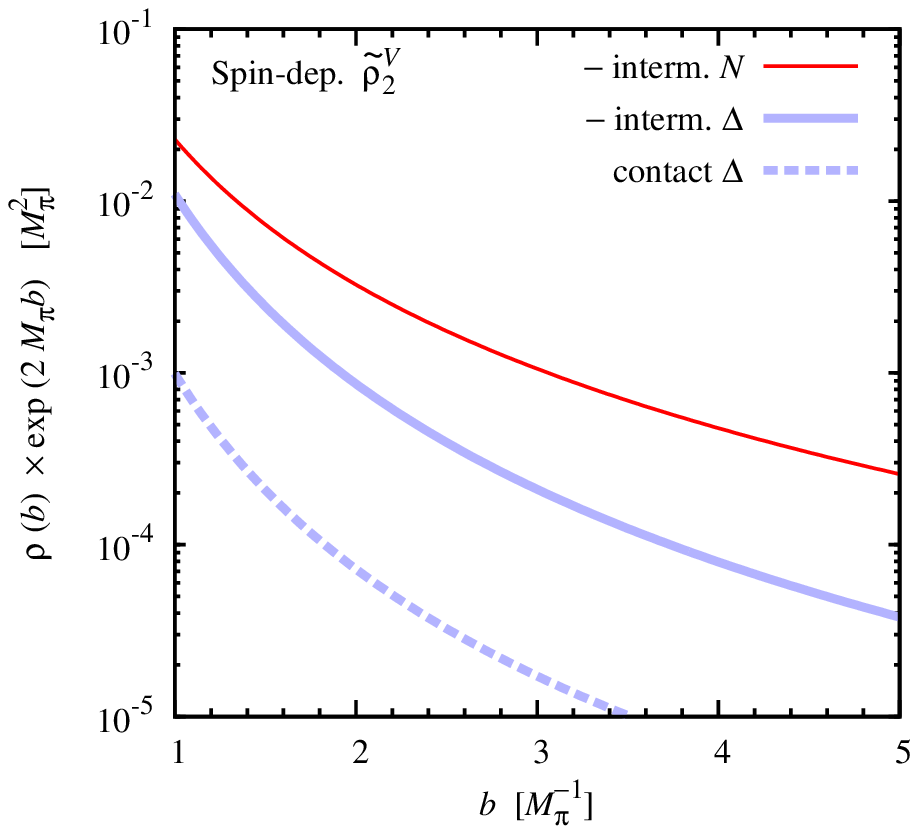}
\\[-4ex]
{\bf (b)}
\end{tabular}
\caption[]{Intermediate-baryon and contact term contributions to the peripheral 
transverse densities.}
\label{fig:contact}
\end{figure}
Numerical results for the contact term densities are shown in Fig.~\ref{fig:contact}a and b.
Their dependence on the distance $b$ is similar to that of the intermediate-baryon contributions 
in the region of interest. The magnitude and sign of the densities follow the pattern implied 
by the large-$N_c$ limit.

The findings regarding the $N_c$-scaling of the contact terms have implications for the 
off-shell behavior of the EFT including isobars (cf.\ Sec.~\ref{sec:chiral}). In the present calculation
we take into account only contact terms resulting from the off-shell behavior of the $\Delta$
propagator and vertices, but do not include any explicit ``new'' $\pi\pi NN$ contact terms associated 
with the introduction of the $\Delta$. Our results show that this prescription gives the correct $N_c$-scaling 
of the transverse densities. From the perspective of the large-$N_c$ limit there is thus {\em no need}
to introduce explicit contact terms. This is positive, as it allows us to perform a consistent calculation with
the minimal off-shell extension. It is also negative, as it means that $N_c$-scaling alone is not sufficient 
to fix the off-shell ambiguity of chiral EFT with isobars. Answers to this question have to come from dynamical 
considerations beyond the large-$N_c$ limit \cite{alarcon16:inprep}.
In this sense the numerical results for the contact term densities should be regarded as estimates, as they 
may be modified by $1/N_c$-suppressed off-shell terms, which we cannot determine within the present scheme.

An advantage of the light-front representation of chiral EFT with isobars is that the effects of the 
off-shell ambiguity are entirely contained in the contact term contributions to the densities. 
The wave function overlap contribution (as defined in the scheme of Sec.~\ref{subsec:current}) 
relies only on the on-shell properties of the EFT. As such this representation is very efficient and might be 
of more general use in studying the off-shell ambiguities of chiral EFT with high-spin particles.
\section{Summary and outlook}
\label{sec:summary}
In this work we have included the $\Delta$ isobar in the light-front representation of chiral EFT
and studied its effect on the nucleon's peripheral transverse densities. We have done this by 
systematically rewriting of the relativistically invariant chiral EFT results in a form that they 
correspond to $N \rightarrow \pi B \; (B = N, \Delta)$ light-front wave function overlap and
contact terms. The resulting expressions give rise to a simple quantum-mechanical picture,
expose the role of pion orbital angular momentum in chiral processes, and illustrate the 
essentially relativistic nature of chiral dynamics. They also permit a concise derivation of the 
$N_c$-scaling relations and numerical evaluation of the peripheral densities. It is worth
emphasizing that (a) the light-front representation presented here is {\it exactly equivalent} 
to the relativistically invariant chiral EFT results, which implies that rotational invariance is
effectively preserved in our approach; (b) the quantum-mechanical picture described here is based on 
chiral EFT interactions and has an objective physical meaning, which distinguishes it from 
phenomenological pion cloud models or other models of transverse nucleon structure.

The $\Delta$ isobar affects the transverse densities both parametrically and numerically.
Parametrically it restores the proper scaling behavior of the peripheral densities in the
large-$N_c$ limit. The scaling behavior involves both the light-front wave function overlap 
and the contact terms, and provides a non-trivial test of the formalism.
Numerically we find that the intermediate isobar contributions are of moderate size in the 
spin-dependent and -independent densities, $\rho_1(b)$ and $\widetilde\rho_2(b)$,
amounting to $\sim$30\%--10\% of the intermediate nucleon contributions at
$b$ = 1--5 $M_\pi^{-1}$. The isobar contributions become very substantial in the right current 
density $\rho_{\rm right} = \rho_1 + \widetilde\rho_2$, in which the nucleon contribution 
is small, and inclusion of the isobar reverses the sign of the density at $b \lesssim 3\, M_\pi^{-1}$.

The practical issues in extracting peripheral transverse densities from elastic form factor 
data and testing the chiral predictions have been discussed in
Refs.~\cite{Strikman:2010pu,Granados:2013moa,Granados:2015rra}. Because the analytic properties of
the form factors (especially the two-pion cut at $t > 4 M_\pi^2$) are critically important
in the Fourier transform at large distances, one must use dispersive fits to the form factor data 
when extracting the peripheral densities. A dispersive analysis of the transverse densities has shown
that the chiral component becomes dominant at $b \gtrsim 2\, \textrm{fm} \sim 1.4 \, M_\pi^{-1}$;
at smaller distances the densities are dominated by the $\rho$ meson resonance \cite{Miller:2011du}.
A detailed phenomenological study of the uncertainties of the peripheral densities, using dispersive 
fits to present form factor data and including their theoretical uncertainties, should be undertaken
in the future.

The theoretical methods developed in this study could be applied to several other problems of interest.
The derivation of the light-front representation from invariant integrals in Secs.~\ref{subsec:current}
and \ref{subsec:overlap} represents a general scheme that could be applied to EFTs with other high-spin 
particles, or to matrix elements of higher-spin operators such as those describing moments of
generalized parton distributions (twist-2, spin-$n$ operators, $n \geq 1$). 
It should generally be useful in situations where rotational invariance is critical 
(e.g., to avoid power-like divergences) and a relativistically invariant
formulation is available. One obvious example are the nucleon form factors of the energy-momentum
tensor (twist-2, spin-2 operator), which describe the distribution of momentum, mass, and forces in 
the nucleon and are an object of intense study in connection with nucleon spin 
problem; see Ref.~\cite{Leader:2013jra} for a review.

In the present study we calculate the transverse densities in the nucleon and consider the isobar
as an intermediate state in the peripheral chiral processes. The same formalism would allow one
to calculate also the transverse densities associated with the $N \rightarrow \Delta$ transition
form factors, for which chiral EFT results 
and experimental data are available \cite{Pascalutsa:2005ts,Pascalutsa:2007wz,Carlson:2007xd}. 
The peripheral current matrix element between the
spin-1/2 and spin-3/2 states contain structures with orbital angular momentum $\Delta L = 0, 1$ and 2, 
which could be illustrated in the mechanical picture of Sec.~\ref{sec:mechanical}. The formalism could be 
applied just as well to the transverse densities in the isobar state itself, associated with the 
$\Delta \rightarrow \Delta$ form factors, which are a well-defined concept in the context of 
chiral EFT \cite{Ledwig:2010ya,Ledwig:2011cx}, 
and are studied also in Lattice QCD \cite{Alexandrou:2009hs,Alexandrou:2008bn,Aubin:2008qp}. 
\acknowledgments
We thank J.~M.~Alarcon and J.~Goity for helpful discussions of questions regarding chiral EFT and the
$\Delta$ isobar.

This material is based upon work supported by the U.S.~Department of Energy, Office of Science, 
Office of Nuclear Physics under contract DE-AC05-06OR23177.
\end{document}